\RequirePackage{fix-cm}
\RequirePackage{amsmath,amssymb}
\documentclass[twocolumn,epjc3]{my_svjour3}  
\smartqed  
\RequirePackage{mathptmx}
\RequirePackage{latexsym}
\RequirePackage[numbers,sort&compress]{natbib}
\RequirePackage[colorlinks,citecolor=blue,urlcolor=blue,linkcolor=blue]{hyperref}
\RequirePackage{graphicx}
\usepackage[utf8]{inputenc}
\usepackage{tabularx}
\usepackage{mathtools}
\usepackage{longtable}
\usepackage{times}
\usepackage{subcaption} 
\usepackage{enumerate}

\usepackage{microtype}
\usepackage{lmodern}
\usepackage[outercaption]{sidecap}
\usepackage{macros}
\usepackage{xspace}
\usepackage{cancel}
\usepackage{array}
\usepackage{siunitx}
\usepackage{lscape}
\usepackage{booktabs}
\usepackage{macros}

\usepackage{color}
\usepackage{xcolor}

\usepackage{etoolbox}
\usepackage[absolute,overlay]{textpos}
\makeatletter
\def\@mkboth#1#2{}
\newlength\appendixwidth
\preto\appendix{\addtocontents{toc}{\protect\patchl@section}}
\newcommand{\patchl@section}{%
	\settowidth{\appendixwidth}{\textbf{Appendix }}%
	\addtolength{\appendixwidth}{1.5em}%
	\patchcmd{\l@section}{1.5em}{\appendixwidth}{}{\ddt}%
}
\makeatother

\journalname{Eur. Phys. J. C}

\begin{document}

\title{Ward identity determination of $Z_\mathrm{S}/Z_\mathrm{P}$ for $N_\mathrm{f}=3$ lattice QCD in a Schrödinger functional setup
}

\subtitle{
	\\
	\includegraphics[width=2.5cm]{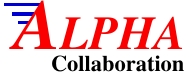}
} 

\author{
	Jochen Heitger\thanksref{addr1}
        \and
        Fabian Joswig\thanksref{e1,addr1}
        \and
        Anastassios Vladikas\thanksref{addr2}
}
\thankstext{e1}{e-mail: fabian.joswig@wwu.de}

\institute{Westf\"alische Wilhelms-Universit\"at M\"unster, Institut f\"ur Theoretische Physik, Wilhelm-Klemm-Stra{\ss}e 9, 48149 M\"unster, Germany \label{addr1}
           \and
           INFN, ``Rome Tor Vergata'' Division, c/o Dipartimento di Fisica, Via della Ricerca Scientifica~1, 00133 Rome, Italy \label{addr2}\\
}

\date{May 4, 2020}

\maketitle

\begin{textblock*}{3cm}(16.25cm,3.2cm) 

	MS-TP-20-19

\end{textblock*}

\begin{abstract}
We derive chiral Ward identities for lattice QCD with Wilson quarks and $\NF \ge 3$ flavours, on small lattices with 
Schrödinger functional boundary conditions and vanishingly small quark masses. These identities relate the axial variation of the
non-singlet pseudoscalar density to the scalar one, thus enabling the non-perturbative determination of the scale-independent 
ratio $Z_\mathrm{S}/Z_\mathrm{P}$ of the renormalisation parameters of these operators. We obtain results for $\Nf=3$
QCD with tree-level Symanzik-improved gluons and Wilson-Clover quarks, for bare gauge couplings which cover the typical range of 
large-volume $\Nf = 2+1$ simulations with Wilson fermions at lattice spacings below $0.1\,$fm. The precision of our results varies
from 0.3\% to 1\%, except for the coarsest lattice, where it is 2\%. We discuss how the $Z_\mathrm{S}/Z_\mathrm{P}$ ratio can be used in the non-perturbative calculations of 
$\mathrm{O}(a)$ improved renormalised quark masses.
\end{abstract}

\tableofcontents

\section{Introduction}

Lattice QCD with Wilson fermions is a long-established regularisation. The fermionic action satisfies most desirable properties, namely strict locality, lack of fermion doublers, and preservation of flavour symmetry in a straightforward way.  Well-known shortcomings are the presence of discretisation effects linear in the lattice spacing and, most importantly, the loss of chiral symmetry. The first problem is solved by applying the Symanzik-improvement programme (see for instance Ref.~\cite{Luscher:1998pe} for a review and Ref.~\cite{Luscher:1996sc} for more details). Chiral symmetry is recovered in the continuum, at the cost of having to deal with complicated renormalisation properties for most quantities of interest (cf. Ref.~\cite{Bochicchio:1985xa} and references therein; for a review see also Ref.~\cite{Vladikas:2011bp}). A frequently cited example of these complications is the power divergence $\m_{\rm crit} \sim 1/a$, which must be subtracted from bare quark masses before they are renormalised multiplicatively. Other examples are the normalisation parameter $Z_\mathrm{A}$ of the axial current and the ratio $Z_\mathrm{S}/Z_\mathrm{P}$ of the non-singlet scalar and pseudoscalar density renormalisation parameters. In a regularisation scheme which respects chiral symmetry, these quantities are strictly equal to unity at finite values of the UV cutoff. With Wilson fermions these quantities are scale-independent finite functions of the gauge coupling, which tend to unity as we approach the continuum limit.  In principle they are determined by requiring that chiral Ward identities at non-vanishing lattice spacing tend to their formal counter-parts in the continuum limit.  The scope of this paper is to provide a method for the determination of $Z_\mathrm {S}/Z_\mathrm{P}$ based on Ward identities on physically
small lattices with Schrödinger functional boundary conditions and realising a line of constant physics (LCP) in parameter space. Results are obtained for $\NF=3$ dynamical quarks.

The general idea behind using chiral Ward identities in order to evaluate $Z_\mathrm {S}/Z_\mathrm{P}$ for Wilson fermions appeared in Ref.~\cite{Bochicchio:1985xa}\footnote{In practice, distinct chiral Ward identities are used for the computation of the ratio $Z_\mathrm{S}/(Z_\mathrm{P} Z_\mathrm{A})$ and $Z_\mathrm{A}$; the two results are subsequently multiplied to give $Z_\mathrm {S}/Z_\mathrm{P}$.}. It has been put to practice with quenched, unimproved Wilson fermions in Ref.~\cite{Maiani:1987by} and subsequently with tree-level Symanzik-improved ones in Ref.~\cite{Martinelli:1993dq}. The chiral Ward identities in question were obtained for large-volume lattices with periodic boundary conditions and non-chiral quark masses. Ratios of $Z_\mathrm {S}/Z_\mathrm{P}$ were calculated at fixed gauge coupling for several quark masses and extrapolated to the chiral limit. A second-generation of calculations was not based on Ward identities but obtained by computing $Z_\mathrm {S}$ and $Z_\mathrm{P}$ in the RI/MOM scheme~\cite{Martinelli:1994ty}. Again these calculations are performed at finite quark masses, followed by chiral extrapolations. A well known problem in this approach is that the $Z_\mathrm {S}/Z_\mathrm{P}$ ratio thus obtained differs from the Ward identity one by "Goldstone pole contaminations" at the IR end of a renormalisation window. This problem was first identified in Ref.~\cite{Martinelli:1994ty}, and subsequently discussed in Refs.~\cite{Cudell:1998ic,Cudell:2001ny,Giusti:2000jr,Papinutto:thesis} (and reviewed in Ref.~\cite{Vladikas:2011bp}), while the discussion specific to the difference between Ward identity and RI/MOM determinations of the ratio $Z_\mathrm {S}/Z_\mathrm{P}$ is found in Ref.~\cite{Giusti:2000jr}. Although the problem is greatly attenuated by the RI/SMOM variant of this method~\cite{Sturm:2009kb},  the requirement of a reliable renormalisation window is inherent in these approaches.

In the present work we revisit the Ward identity method, with an important novelty: lattices with small physical volumes and Schrödinger functional boundary conditions are used, with quark flavours degenerate in mass and (almost) at the chiral limit. In doing so, we follow closely the method introduced in Ref.~\cite{Luscher:1996jn} (and
originally applied in the quenched approximation in that work) for the non-perturbative determination of the scale independent normalisation parameter $Z_\mathrm{A}$ of the axial vector current. Updates and optimisations of these computations can be found in refs.~\cite{DellaMorte:2005rd,Bulava:2016ktf} for two- and three-flavour QCD, respectively. Ward identities are imposed at constant physics to ensure a removal of $\mathrm{O}(a)$ effects in on-shell quantities and, at the same time, smoothly vanishing $\mathrm{O}(a^2)$ effects as the bare coupling is varied. It must be stressed that the chiral Ward identities adopted in these works to determine $Z_\mathrm{A}$ are valid for $\NF \geq 2$ quark flavours, while the ones we introduce in the present work for the determination of $Z_\mathrm{S}/(Z_\mathrm{P} Z_\mathrm{A})$ are valid for $\NF \geq 3$.

We note in passing that, based on the chirally rotated Schr\"odinger functional construction of Ref.~\cite{Sint:2010eh}, a more recent method for the non-perturbative computation of $Z_\mathrm{S}/Z_\mathrm{P}$ has been mentioned in Ref.~\cite{Brida:2016rmy}.

This paper is organised as follows: in Section~\ref{sec:ward_identities} (Subsection~\ref{subsec:ward_identities_formal}) we formally derive chiral Ward identities for continuum QCD, which relate correlation functions of non-singlet pseudoscalar and scalar composite operators (densities). The former are correlation functions with two operator insertions at two distinct space-time points (an axial current and a pseudoscalar density) in the presence of a generic external source operator. The latter involve a single insertion of the scalar operator. Subsequently (Subsection~\ref{subsec:ward_identities_SF}), we rewrite the same Ward identities in the lattice-regularised QCD with Wilson fermions. The external source consists of two standard Schr\"odinger functional boundary sources, each placed at a temporal boundary. The loss of chiral symmetry by Wilson fermions is taken into account by the renormalisation constants $Z_\mathrm {P}$ and $Z_\mathrm {S}$ of the pseudoscalar and scalar densities and the normalisation of the axial current, $Z_\mathrm {A}$. In the chiral limit, these Ward identities hold up to $\rmO(a^2)$ discretisation effects. We also discuss the corrections arising in practical simulations, which slightly deviate from the chiral limit; these are  $\rmO(am,a^2)$. Finally, in Subsection~\ref{subsec:ward_identities_Wick} we re-express these Ward identities in terms of traces of valence quark propagators, which multiply factored-out traces of generators of the $SU(\NF)$ flavour group.

Section~\ref{sec:corr-funct} takes an even closer look at these Ward identities. We distinguish several equivalence classes, each consisting of identities with different flavour structure, which reduce to the same relations between correlation functions, giving the same $Z_\mathrm{S}/(Z_\mathrm{P} Z_\mathrm{A})$ result. Ward identities belonging to different equivalence classes provide $Z_\mathrm{S}/(Z_\mathrm{P} Z_\mathrm{A})$ estimates which differ by $\rmO(am,a^2)$ effects. If we neglect these effects, we can combine identities from different equivalence classes, ending up with new relations between correlation functions (true up to $\rmO(am,a^2)$ errors). Thus we can explore to what extent different equivalence classes provide independent estimates of $Z_\mathrm{S}/(Z_\mathrm{P} Z_\mathrm{A})$. Some of these estimates are expected to be noisier than others, as they are obtained using both quark-connected and quark-disconnected correlation functions.

In Section~\ref{section:results} we present our results for QCD with $\NF = 3$ dynamical flavours,
where the lattice gauge action is tree-level Symanzik-improved and the fermion action is non-perturbatively Wilson-Clover improved. Our simulations are performed with degenerate mass flavours lying close to the chiral limit. The non-perturbative determination of the ratio $Z_\mathrm{S}/Z_\mathrm{P}$ is carried out along a line of constant physics in parameter space. In practice, this requirement is met by ensuring a volume of almost constant spatial extent $L\sim1.2\,$fm in physical units, with Schrödinger functional boundary conditions. The ratio between temporal and spatial extent $T/L$ is also kept fixed. This implies that any remaining intrinsic ambiguities in $Z_\mathrm{S}/Z_\mathrm{P}$ of $\rmO(a^2)$ or higher (in the $\rmO(a)$ improved setup adopted here) disappear smoothly towards the continuum limit. 
The gauge couplings of our simulations span a range typical for the computations performed by the CLS (Coordinated Lattice Simulations) effort in QCD with $\NF=2+1$ flavours of non-perturbatively improved Wilson fermions~\cite{Bruno:2014jqa,Bruno:2016plf,Bali:2016umi,Mohler:2017wnb}. Our $Z_\mathrm{S}/(Z_\mathrm{P} Z_\mathrm{A})$ results are divided out by $Z_\mathrm{A}$, estimated in Ref.~\cite{DallaBrida:2018tpn}. Our $Z_\mathrm{S}/Z_\mathrm{P}$ estimates are subsequently extrapolated to the chiral limit at fixed $g_0^2$. Results are obtained from several Ward identities; they differ by discretisation effects. Thus it is possible to create ratios of the different $Z_\mathrm{S}/Z_\mathrm{P}$ determinations, and plot them against (powers of) the lattice spacing, confirming the
expected scaling behaviour. The statistically and systematically most
precise $Z_\mathrm{S}/Z_\mathrm{P}$ determination is parameterised as a continuous function of $g_0^2$, which is our final answer. 
This is compared to two other determinations: one is based on ratios of PCAC quark masses with different flavours, employing essentially the same small-volume Schrödinger functional setup~\cite{deDivitiis:2019xla}; the other is based on the relation between bare current quark masses and bare subtracted quark masses, computed on large volumes with open boundary conditions~\cite{Bali:2016umi}.

Finally, in Section~\ref{section:qmasses} we discuss how $Z_\mathrm{S}/Z_\mathrm{P}$ can be used in quark mass determinations along the lines proposed in Ref.~\cite{Durr:2010aw}, but performing the mass renormalisation in the Schrödinger functional scheme and the renormalisation group running non-perturbatively, between renormalisation scales $\mu_{\mathrm{had}} \sim \Lambda_{\mathrm{QCD}}$ and $\mu_\mathrm{PT} \sim M_{\mathrm W}$. Such a calculation is subjected to different systematics than the standard ALPHA-CLS method, recently applied in Ref.~\cite{Bruno:2019vup}.
 
Work in progress culminating to this paper had been reported in Refs.~\cite{Heitger:2017njs,Heitger:2018pwb}.

\section{Chiral Ward identities for \texorpdfstring{$Z_\mathrm{S}/Z_\mathrm{P}$}{ZS/ZP}}
\label{sec:ward_identities}

In this Section we will derive chiral Ward identities which relate correlation functions of non-singlet scalar and pseudoscalar composite operators (densities).
These enable us to compute non-perturbatively the ratio $Z_\mathrm{S}/Z_\mathrm{P}$, which determines the relative normalisation of these
scalar and pseudoscalar densities when the regularisation (Wilson fermion action) breaks chiral symmetry. 
First we will derive the pertinent chiral Ward identities in the formal continuum theory. Subsequently, we will show their lattice analogues with Schr\"odinger functional boundary conditions. The resulting Ward identity computation of $Z_\mathrm{S}/Z_\mathrm{P}$ follows very closely that of $Z_\mathrm{A}$, described in refs.~\cite{Luscher:1996jn,DellaMorte:2005rd,Bulava:2016ktf}.

Our notation is pretty standard. Definitions of composite operators of dimension-3, axial transformations and Schr\"odinger functional (SF) boundary
operators are collected in \ref{app:general}. Conventions concerning the $su(\NF)$ flavour algebra are
to be found in \ref{app:sun}. The lattice spacing is denoted by $a$, the (squared) gauge coupling by $g_0^2$, and the inverse lattice coupling by $\beta \equiv 6/g_0^2$. Bare current (PCAC) and subtracted masses are defined in \ref{app:renimp}.

\subsection{Formal chiral Ward identities in the continuum}
\label{subsec:ward_identities_formal}
Under the small axial variations~(\ref{eq:axialtransinft}) of the fermion fields the formal, continuum QCD action in Euclidean space-time transforms as follows:

{\small
\begin{align}
\delta_\mathrm{A} S  = &   \int d^4x \Big [ (\partial_\mu \epsilon^a(x))  A_\mu^a(x)
+ \mathrm{i}  \epsilon^a(x) \bar \psi(x) \{ T^a, M \} \gamma_5 \psi(x) \Big ] \nonumber\\
= & \int d^4x \,\, \epsilon^a(x) \,\, \Big [ - \partial_\mu A_\mu^a(x) +  2m P^a(x) \Big ] \,.
\label{eq:deltaS}
\end{align}
}The fermion mass matrix is denoted by $M$. We work in the flavour symmetric (isospin) limit, so all quark masses $m$ are degenerate. In the last expression we have integrated by parts the term with the axial current.
Chiral Ward identities are obtained by considering that under the change of field variables defined in Eqs.~(\ref{eq:axialtransf}), the expectation value of any composite operator ${\cal O}$ (and products of them) is invariant. In the limit of small axial variations this leads to:
\begin{align}
\delta_\mathrm{A} \langle \cO \rangle  &=  \dfrac{1}{\cal Z} \, \delta_\mathrm{A} \,\, \langle \,\, \int [{\cal D} \psi] [{\cal D} \bar \psi] [{\cal D} G_\mu] \,\,  {\cal O} \,\, \exp(-S) \,\, \rangle =  0 \nonumber \\
\Rightarrow \,\,  \langle \delta_\mathrm{A} {\cal O} \rangle  &=  \langle \cO \,\, \delta_\mathrm{A} S \rangle \,.
\label{eq:genAWI}
\end{align}
We now take the axial variations to be non zero only in a space-time region $R$ with a smooth boundary $\partial R$ (i.e., for $x \in R$, $\epsilon^a(x) \ne 0$; otherwise $\epsilon^a(x) = 0$). The above expression reduces to
\begin{align}
\begin{split}
\int_R d^4x \epsilon^a(x)  \Big [  \partial_\mu \langle A_\mu^a(x) \,\, \cO \rangle - 2m \langle P^a(x) \,\, {\cal O} \rangle \Big ] 
 = - \, \langle \delta_\mathrm{A} {\cal O} \rangle \,.
\end{split}
\end{align}
We consider a product of composite operators $\cO = P^b(y) \cO_{\rm ext}$, where $y \in R$ and $ \cO_{\rm ext}$ is defined outside the region $R$. This implies that $\delta_\mathrm{A} \cO = [\delta_\mathrm{A} P^b(y)] \cO_{\rm ext}$.
The pseudoscalar density  $P^b(x)$ transforms as follows:
\begin{eqnarray}
\delta_\mathrm{A} P^b(x) = \epsilon^c(x) d^{cbe} S^e(x) +  \epsilon^c(x) \dfrac{\delta^{cb}}{\NF} \bar \psi(x) \psi(x) \,.
\label{eq:deltaP}
\end{eqnarray}
At this stage we impose that $\epsilon^c(x) =  \epsilon \delta^{ac}$; i.e., it is a constant phase $\epsilon$ in a fixed direction $a$ in flavour space, so that Ward identities become expressions reflecting {\it global} chiral symmetry. Moreover, in order to sidestep a number of complications\footnote{With Wilson fermions, the singlet scalar operator $\bar \psi(x) \psi(x)$ mixes with the identity operator, introducing the complication of power divergences. Moreover, Wick contractions of the fermion fields of this operator generate quark-disconnected diagrams.},
we chose $a \ne b$, so that the last term on the r.h.s.\ of Eq.~(\ref{eq:deltaP}) drops out\footnote{Here we are working with the algebra $su(\NF)$ for $\NF \ge 3$; for $\NF = 2$ we have that $d^{abe} = 0$ and the r.h.s.\ of Eq.~(\ref{eq:deltaP}) is trivial.}.
Putting everything together, we obtain
\begin{align}
&\int_R d^4x \, \Big [ \partial_\mu \, \langle A_\mu^a(x) P^b(y) \cO_{\rm ext} \rangle - 2m \, \langle P^a(x) P^b(y) \cO_{\rm ext} \rangle \Big ] \nonumber \\
& =  - d^{abe} \langle S^e(y) \cO_{\rm ext} \rangle \,.
\end{align}
We note in passing that the first term is a surface term:

{\small
\begin{align}
\begin{split}
&\int_R d^4x \, \partial_\mu \, \langle A_\mu^a(x) P^b(y) \cO_{\rm ext} \rangle
=\int_{\partial R} d\sigma_\mu(x) \langle A_\mu^a(x) P^b(y) \cO_{\rm ext} \rangle \,.
\end{split}
\end{align}
}As done in  Ref.~\cite{Luscher:1996jn} for $Z_\mathrm{A}$, we chose the region $R$ to be the space-time volume between the hyper-planes at $y_0-t$ and $y_0+t$\footnote{This choice of hyperplanes is made for simplicity. A more general choice, $y_0-t_-$ and $y_0+t_+$, with $t_-\neq t_+$ and $t_-$,$t_+ >0$, is also acceptable.}.
Boundary conditions in space are periodic, implying $\int_R dx_0 d^3 x \partial_k \langle A_k \cdots \rangle = 0$. The Ward identity becomes

{\small
\begin{align}
\begin{split}
& \int d^3{\bf x} \Big \langle \Big [A_0^a(y_0+t;{\bf x}) \, -\, A_0^a(y_0-t;{\bf x}) \Big ] \, P^b(y_0;{\bf y}) \,\, \cO_{\rm ext} \Big \rangle \\
&- 2m \, \int d^3{\bf x} \int_{y_0-t}^{y_0+t} dx_0 \,\, \langle P^a(x_0;{\bf x}) P^b(y_0;{\bf y}) \cO_{\rm ext} \rangle \\
=& - d^{abe} \langle S^e(y) \cO_{\rm ext} \rangle \,.
\end{split}
\end{align}
}It is convenient to introduce a spatial integration over ${\bf y}$:{\small
\begin{eqnarray}
&& \int d^3{\bf y} \int d^3{\bf x} \Big \langle \Big [A_0^a(y_0+t;{\bf x}) - A_0^a(y_0-t;{\bf x}) \Big ] P^b(y_0;{\bf y}) \, \cO_{\rm ext} \Big \rangle \nonumber \\
&&-
2m \, \int d^3{\bf y} \int d^3{\bf x} \int_{y_0-t}^{y_0+t} dx_0 \,\, \langle P^a(x_0;{\bf x}) P^b(y_0;{\bf y}) \cO_{\rm ext} \rangle \nonumber \\
&& = 
- d^{abe} \int d^3{\bf y} \,\, \langle S^e(y) \cO_{\rm ext} \rangle \, .
\label{eq:SPWImassint2}
\end{eqnarray}
}The second line of the l.h.s.\ contains a contact term, arising when $r \equiv |x-y| \to 0$. The operator product is expressed in terms of an OPE (recall that $a \neq b$)
\begin{align}
\begin{split}
P^a(x) \,\, P^b(y)  &\sim d^{abe} \sum_{k=1}^\infty C_k Q^e_{k \, [D]} \, r^{D-6}  \\
&=  d^{abe} C_1  S^e(x) r^{-3}  +  \cdots \,,
\end{split}
\end{align}
where $[D]$ is the operator dimension and the Wilson coefficients $C_k$ contain logarithms. The most divergent term in the OPE, taking into account the various symmetry properties of the operator product, is proportional to $S^e(x)$. The contribution to the space-time volume integral $2m \int_R \cdots$ of a small four-sphere of centre $x$ and radius $a$ (or a small four-cube of size $a$) is then $\sim~m~\int_0^a dr~r^3~r^{D-6} \langle \cdots \rangle \sim m~a^{D-2}~ \langle \cdots \rangle$ and thus the leading term in the OPE contributes $\rmO(am)$. In the lattice regularisation this implies that the contact term contributes an $\rmO(am)$ discretisation effect to the Ward identity, even in a Symanzik-improved setup.

\subsection{Lattice Ward identities with Schrödinger functional boundary conditions}
\label{subsec:ward_identities_SF}
We now adapt the previous formal manipulations to the lattice regularisation with Schr\"odinger functional boundary conditions.
The external source for the Ward identity correlation functions is chosen to be a tensor in flavour space $\cO^{ad}_{\rm ext}$:
\begin{equation}
\cO^{ad}_{\rm ext} = \dfrac{1}{2 L^6} \cO^{\prime a} \cO^d \,,
\label{eq:SFsource2}
\end{equation}
with $\cO^{\prime a}$ and $\cO^d$ defined in Eqs.~(\ref{eq:boundary-sources}).
With this source and in lattice notation the Ward identity~(\ref{eq:SPWImassint2}) becomes (with $b \neq c$):

{\small
\begin{align}
& Z_\mathrm{A} Z_\mathrm{P} \, a^6 \, \times\nonumber \\
& \Bigg \{  \sum_{{\bf x},{\bf y}} \, \langle \cO^{\prime a} \Big [ (A_{\rm I})^b_0(y_0+t; {\bf x}) - (A_{\rm I})^b_0(y_0-t; {\bf x}) \Big ]\, P^c(y_0;{\bf y}) \,  \cO^d \rangle\nonumber \\
& -  2m a \sum_{{\bf x},{\bf y}} \sum_{x_0 = y_0-t}^{y_0+t} w(x_0) \, \langle \cO^{\prime a} P^b(x_0;{\bf x}) \, P^c(y_0;{\bf y}) \,  \cO^d \rangle \Bigg \}\nonumber \\
= & - d^{bce} Z_\mathrm{S} \,\,  a^3 \sum_{\bf y} \langle  \cO^{\prime a}  \, S^e(y) \, \cO^d \rangle  +  \rmO(am,a^2) \,.
\label{eq:SPImassintlatt}
\end{align}
}In this expression, repeated flavour indices $e$ are summed, as usual. The weight factor is $w(x_0) = 1/2$ for $x_0 = y_0 \pm t$ and $w(x_0) = 1$ otherwise. It is introduced in order to implement the trapezoidal rule for discretising integrals. The mass $m$ is the current quark mass defined in Eq.~(\ref{eq:PCACmass}); recall that we work with degenerate masses.

Assuming that we work in the chiral limit (or with nearly vanishing quark masses, so that $\rmO(am)$ effects may be safely neglected),
the above Ward identity is valid up to $\rmO(a^2)$ dicretisation errors in lattice QCD with Wilson quarks. Chiral symmetry breaking implies the (re)normalisation and improvement properties summarised in \ref{app:renimp}. The Symanzik $b$-coefficients appearing in Eqs.~(\ref{eq:A-imp})--(\ref{eq:P-imp}) multiply the subtracted quark mass $m_\mathrm{q}$ or the quark mass matrix $M_\mathrm{q}$. When working in or close to the chiral limit, as is the case in our simulations, we may safely drop these terms. Putting everything together we obtain Ward identity~(\ref{eq:SPImassintlatt}). The renormalisation factors of the external sources $\cO^{\prime a}$ and $\cO^d$ are not taken into consideration, as they cancel out on both sides of the identity.  Note that the term proportional to the current quark mass $m$ may also be dropped in the chiral limit.
In practice, since we are always working with masses that are not strictly zero, it turns out that it is advantageous to keep this term; see Ref.~\cite{Bulava:2016ktf} and Section \ref{sec:chiral_extrapolations}.

Eq.~(\ref{eq:SPImassintlatt}) can be solved for $Z_\mathrm{S}/(Z_\mathrm{P} Z_\mathrm{A})$. With $Z_\mathrm{A}$ known either from other PCAC Ward identities~\cite{Luscher:1996jn,DellaMorte:2005rd,Bulava:2016ktf} or from the chirally rotated Schrödinger functional formalism~\cite{DallaBrida:2018tpn}, we can thus obtain $Z_\mathrm{S}/Z_\mathrm{P}$.

\subsection{Lattice Ward identities, Wick contractions, and fla\-vour factors}
\label{subsec:ward_identities_Wick}
\begin{figure*}[tb]
	\centering
	\begin{subfigure}[b]{0.32\textwidth}
		\centering
		\includegraphics[width=0.85\textwidth]{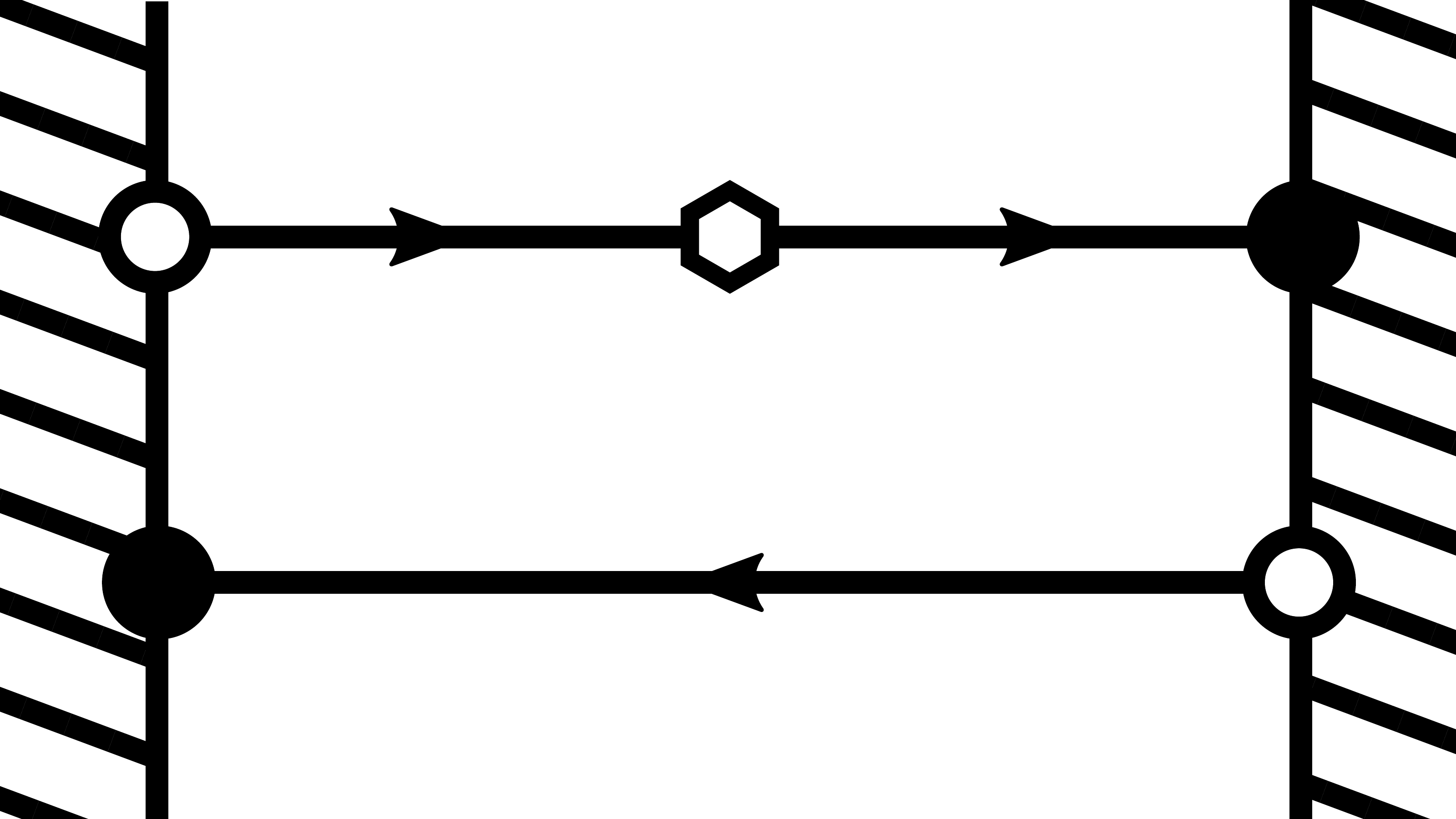}
		\caption{Diagram $ F_{\mathrm{S;1}}$}
	\end{subfigure}
	\begin{subfigure}[b]{0.32\textwidth}
		\centering
		\includegraphics[width=0.85\textwidth]{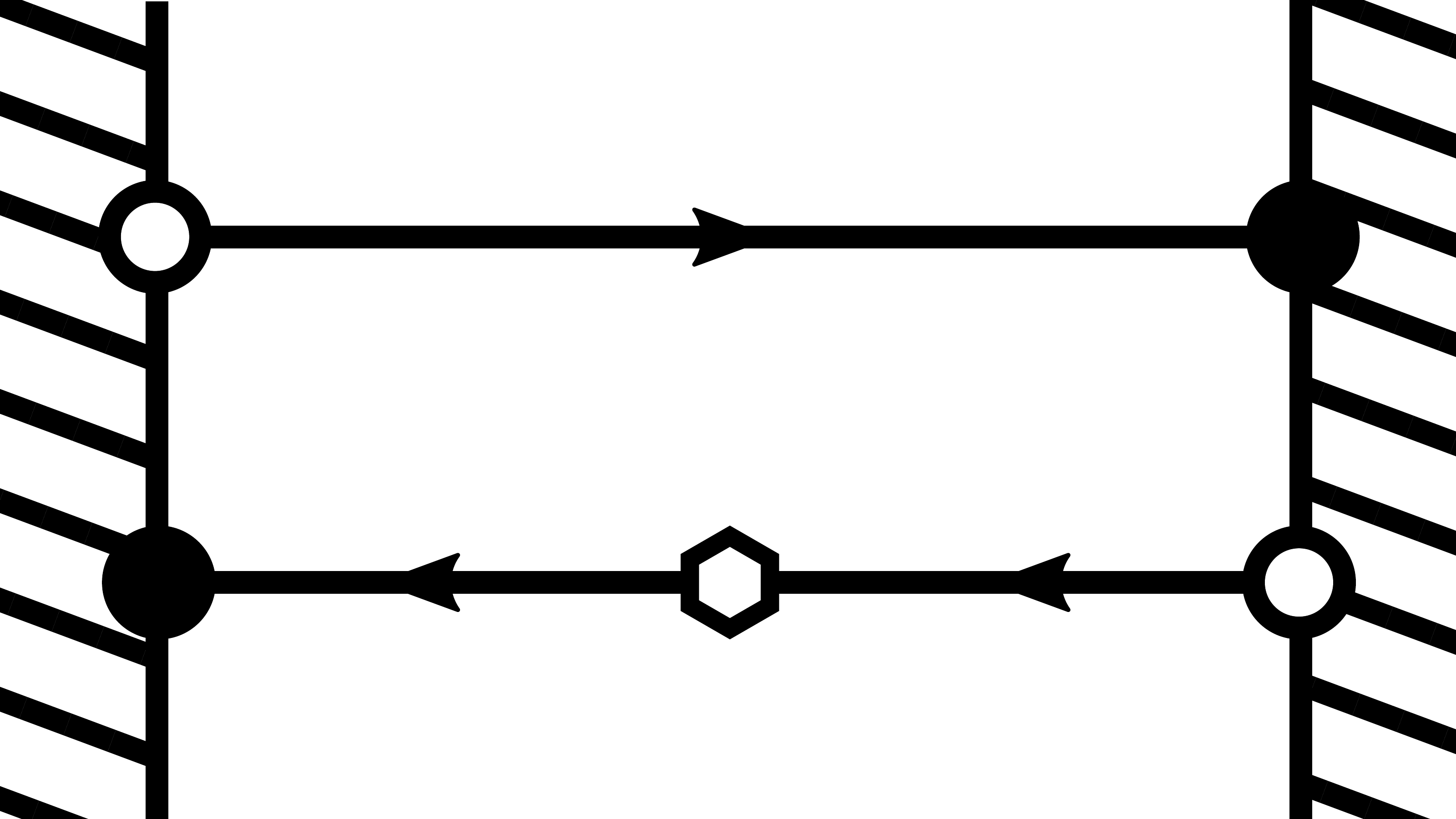}
		\caption{Diagram $ F_{\mathrm{S;2}}$}
	\end{subfigure}
	\begin{subfigure}[b]{0.32\textwidth}
		\centering
		\includegraphics[width=0.85\textwidth]{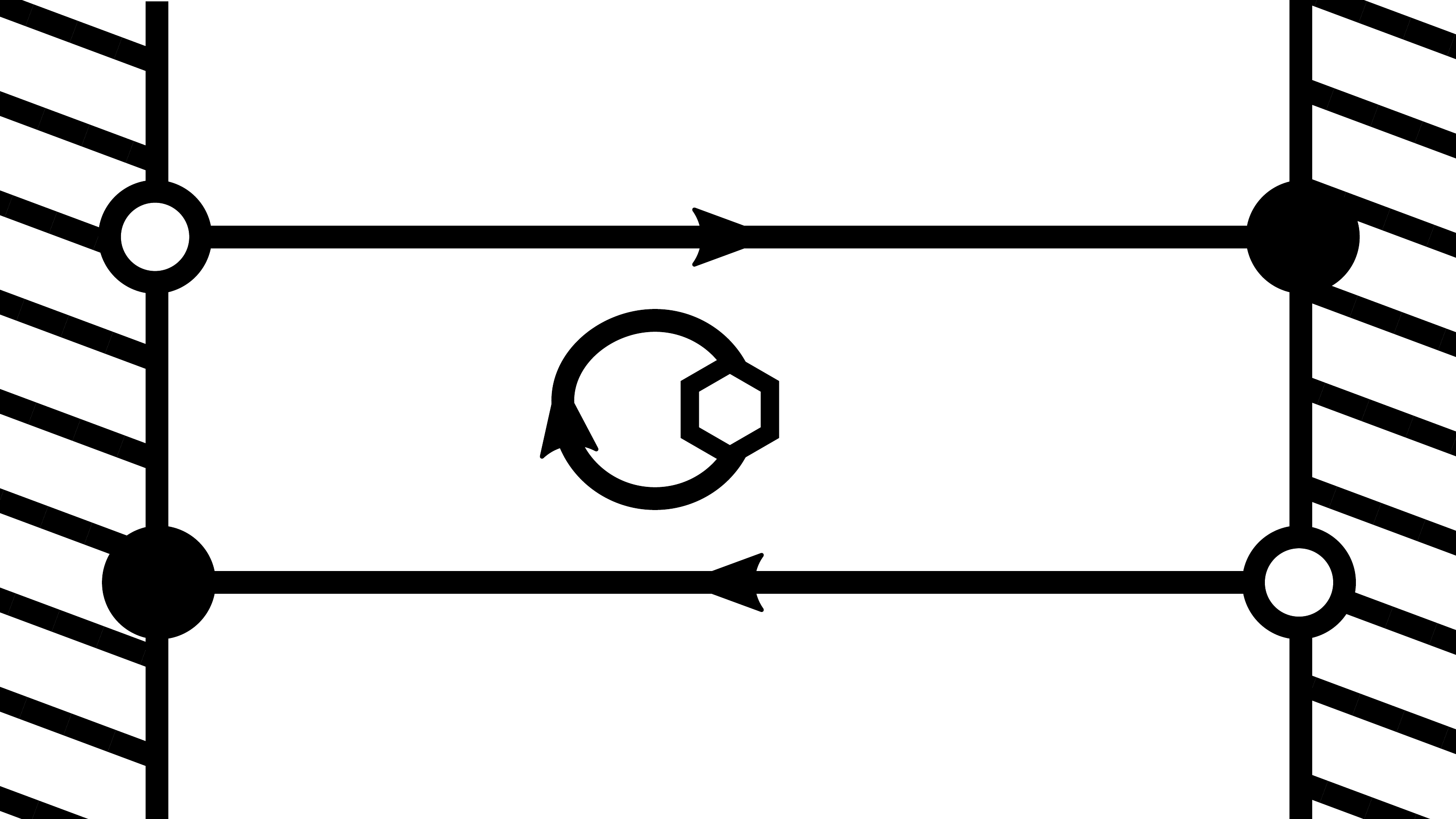}
		\caption{Quark-disconnected diagram}
	\end{subfigure}
	\caption{The trace diagrams contributing to the expectation values of Table~\ref{tab:trBOOB}. The leftmost (rightmost) wall is  time-slice $x_0 = 0$ ($x_0 = T$) with a $\gamma_5$ Dirac matrix between circles. The hexagons in the bulk represent the insertions of a scalar operator $S(y)$. The open circles correspond to the boundary fields $\zeta$ (at $x_0 = 0$) and $\zeta^\prime$ (at $x_0 = T$), while the filled circles denote $\bar \zeta$ (at $x_0 = 0$) and $\bar \zeta^\prime$ (at $x_0 = T$). Quark-connected diagrams $ F_{\mathrm{S;1}}$ and $ F_{\mathrm{S;2}}$ are single traces, formed by starting from any point and following the lines (quark propagators) around until we close the loop. The quark-disconnected diagram is a product of two traces.}
	\label{fig:WI-S}
\end{figure*}
Ward identity~(\ref{eq:SPImassintlatt}) relates expectation values of four composite operators on the l.h.s.\ to those of three composite operators on the r.h.s.; with a slight abuse of terminology, we call these four- and three-point correlation functions, respectively. We express these correlation functions, with Schr\"odinger functional boundary fields, in terms of traces of quark propagators. In standard ALPHA notation \cite{Luscher:1996vw}, $[\psi(y) \, \bar \psi(x)]_\mathrm{F}$ denotes a quark propagator in a fixed background gauge field configuration, where $x$ and $y$ are space-time points in the bulk of the lattice. Propagators from the $x_0=0$ boundary to the bulk are $[\zeta({\bf v}) \bar \psi(y)]_{\mathrm F}$ (with $\bf v$ a point at the $x_0=0$ boundary), while those from the  $x_0=T$ boundary to the bulk are $[\zeta^\prime({\bf v^\prime}) \bar \psi(y)]_{\mathrm F}$ (with $\bf v^\prime$ a point at $x_0=T$). Boundary-to-boundary propagators are  $[\zeta^\prime({\bf v^\prime}) \bar \zeta({\bf u})]_{\mathrm F}$. For proper definitions see Ref.~\cite{Luscher:1996vw}. Note that, since we are working in the $su(\NF)$-symmetric limit, all masses are degenerate and quark propagators of different flavours are indistinguishable\footnote{The notation for fermion fields is somewhat ambiguous: for example, while in this Subsection $\psi(x), \zeta({\bf v}), \zeta^\prime({\bf v^\prime})$ etc. stand for fields of a single flavour, in \ref{app:general} the same quantities denote column vectors in flavour space. This ambiguity is fairly standard and should not create confusion.}.

Performing the Wick contractions, we write the three-point correlation function of Eq.~(\ref{eq:SPImassintlatt}) as
\begin{align}
\begin{split}
\label{eq:OSO=FafaFbfb}
&a^3 \sum_{\bf y}\langle \cO^{\prime a} \, S^e(y) \,\cO^d \rangle \\
 = &- \mathrm{i} a^{15} \Big ( T^{dea} F_{\mathrm{S;1}}(y_0) +T^{aed} F_{\mathrm{S;2}}(y_0) \Big ) \,,
\end{split}
\end{align}
where $T^{aed} \equiv \Tr(T^aT^eT^d)$ are traces of three flavour $su(\NF)$ generators and $F_{\mathrm{S;1}}(y_0),F_{\mathrm{S;2}}(y_0)$ are expectation values of traces of quark propagators with a scalar insertion. The exact expressions can be found in Table~\ref{tab:trBOOB}.
Note that traces $\Tr$ act in flavour space, traces $\tr$ act in spin-colour space, and $\langle \cdots \rangle$ denote averages over gauge field configurations. In Fig.~\ref{fig:WI-S} we show the quark-line diagrams corresponding to the spin-colour traces in the above equation. Any Wick  contraction between fermion fields at the same point in the bulk $[\psi(y), \bar\psi(y)]_{\rm F}$, or between boundary fields at the same time-slice (e.g. $[\zeta({\bf v}) \bar \zeta({\bf u})]_{\rm F}$) gives rise to a quark-disconnected diagram\footnote{It is common practice to refer to these diagrams simply as disconnected. Since from a strict field-theoretic point of view they are connected (with multitudes of gluon lines, some of which contain fermion loops), the term quark-disconnected is more appropriate (valence-quark-disconnected would be even more accurate, but far too long). In the literature, quark-connected and quark-disconnected are sometimes referred to as one- and two-boundary diagrams.}, multiplied by the trace of an $su(\NF)$ generator. As this trace is zero, such diagrams do not contribute to the three-point correlation function. An example of such a diagram is shown in Fig.~\ref{fig:WI-S}.

In \ref{app:corr-funct-symm} we combine the usual $\gamma_5$-Hermiticity property of quark propagators, charge conjugation invariance of the lattice theory, and the trace properties of Eq.~(\ref{eq:trTTT}), to cast the r.h.s.\ of Eq.~(\ref{eq:OSO=FafaFbfb}) into a single real term, and obtain for the r.h.s.\ of the Ward identity~(\ref{eq:SPImassintlatt}):
\begin{align}
\label{eq:WIrhs}
\mathrm{WI~r.h.s.} = - \frac{a^{15}}{2} Z_{\mathrm S}  d^{bce} d^{ade} \Re \Big [ F_{\mathrm{S;1}}(y_0) \Big ] \,.
\end{align}
Next we concentrate on the l.h.s.\ of Eq.~(\ref{eq:SPImassintlatt}). For simplicity we drop, for the moment, the term proportional to the quark mass. The l.h.s.\ consists of boundary-to-boundary correlation functions with two insertions of dimension-3 operators in the bulk, which can be cast in the general form
\begin{figure*}[tb]
	\centering
	\begin{subfigure}[b]{0.32\textwidth}
		\centering
		\includegraphics[width=0.85\textwidth]{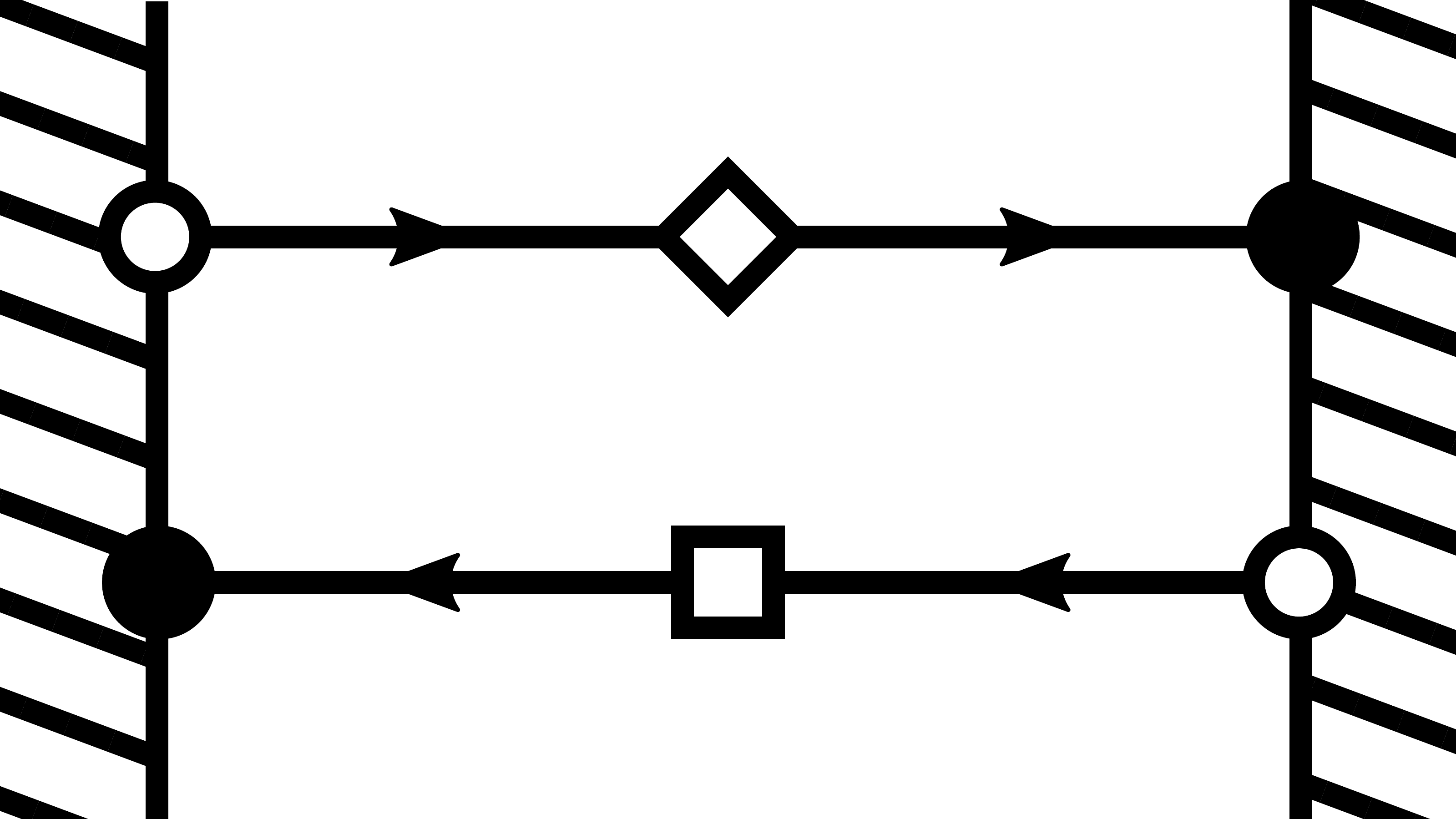}
		\caption{Diagram $F_{\mathrm{AP};3}$}
	\end{subfigure}
	\begin{subfigure}[b]{0.32\textwidth}
		\centering
		\includegraphics[width=0.85\textwidth]{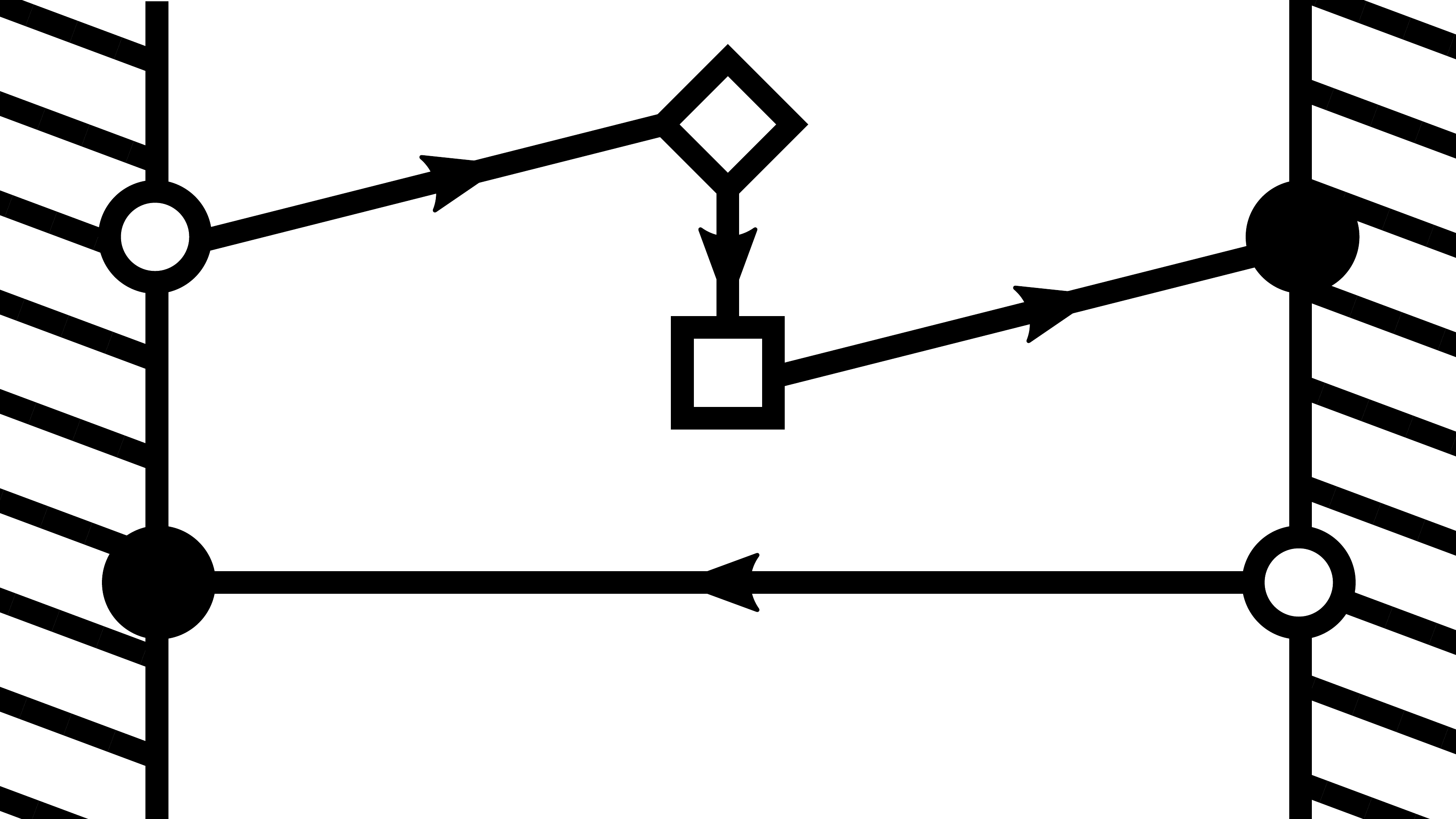}
		\caption{Diagram $F_{\mathrm{AP};5}$}
	\end{subfigure}
	\begin{subfigure}[b]{0.32\textwidth}
		\centering
		\includegraphics[width=0.85\textwidth]{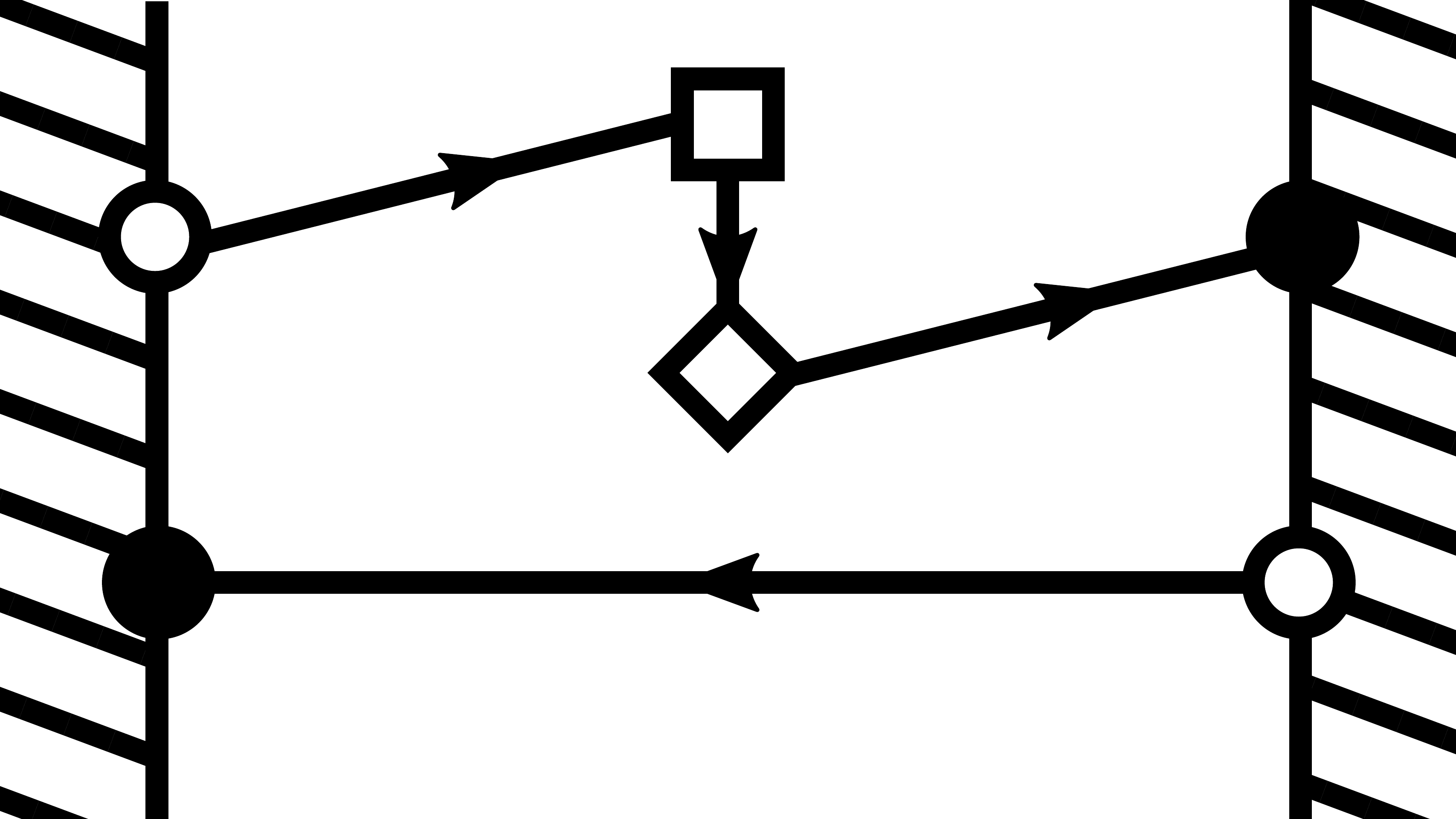}
		\caption{Diagram $F_{\mathrm{AP};1}$}
	\end{subfigure}
	\begin{subfigure}[b]{0.32\textwidth}
		\centering
		\includegraphics[width=0.85\textwidth]{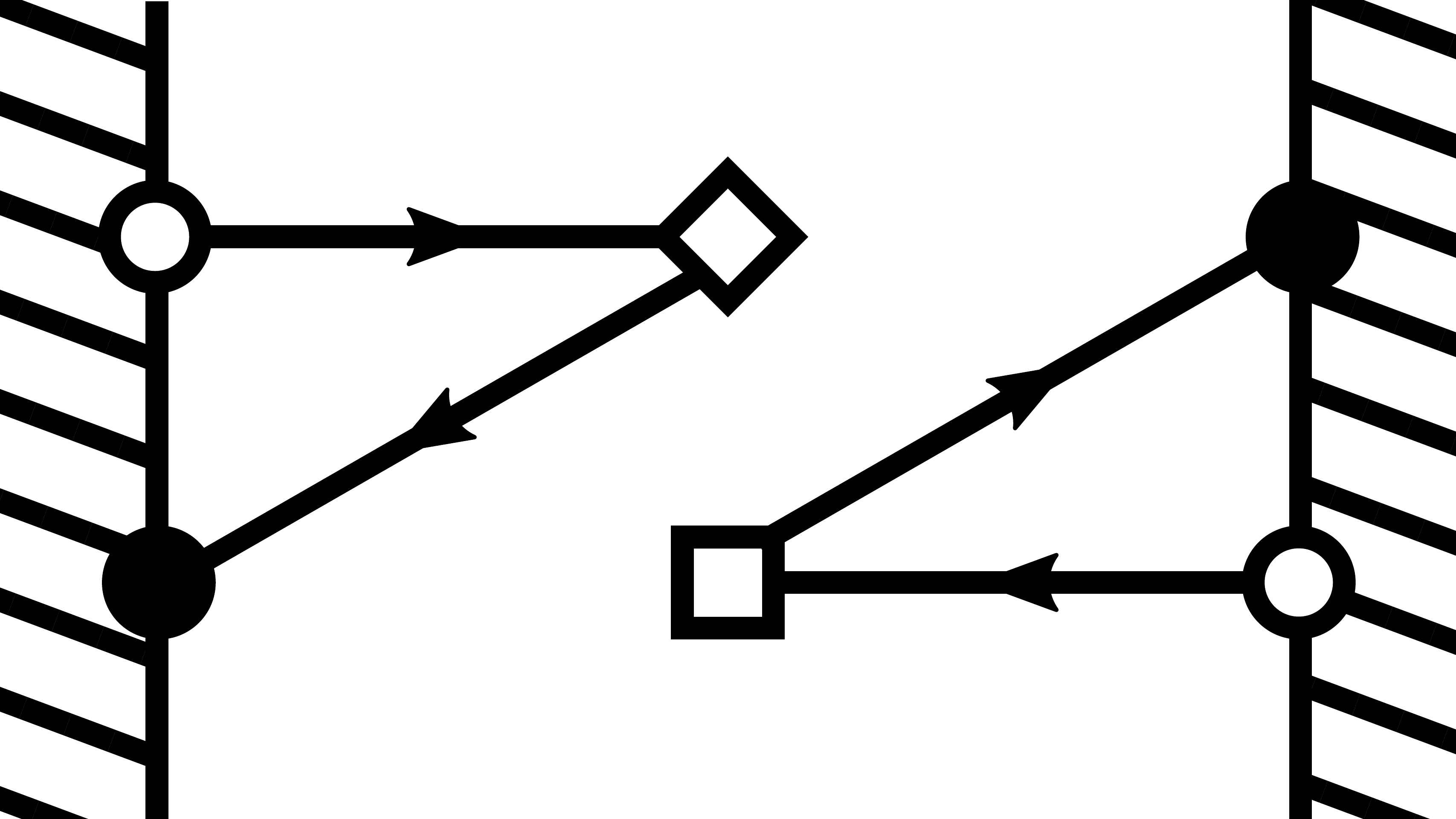}
		\caption{Diagram $F_{\mathrm{AP};8}$}
	\end{subfigure}
	\begin{subfigure}[b]{0.32\textwidth}
		\centering
		\includegraphics[width=0.85\textwidth]{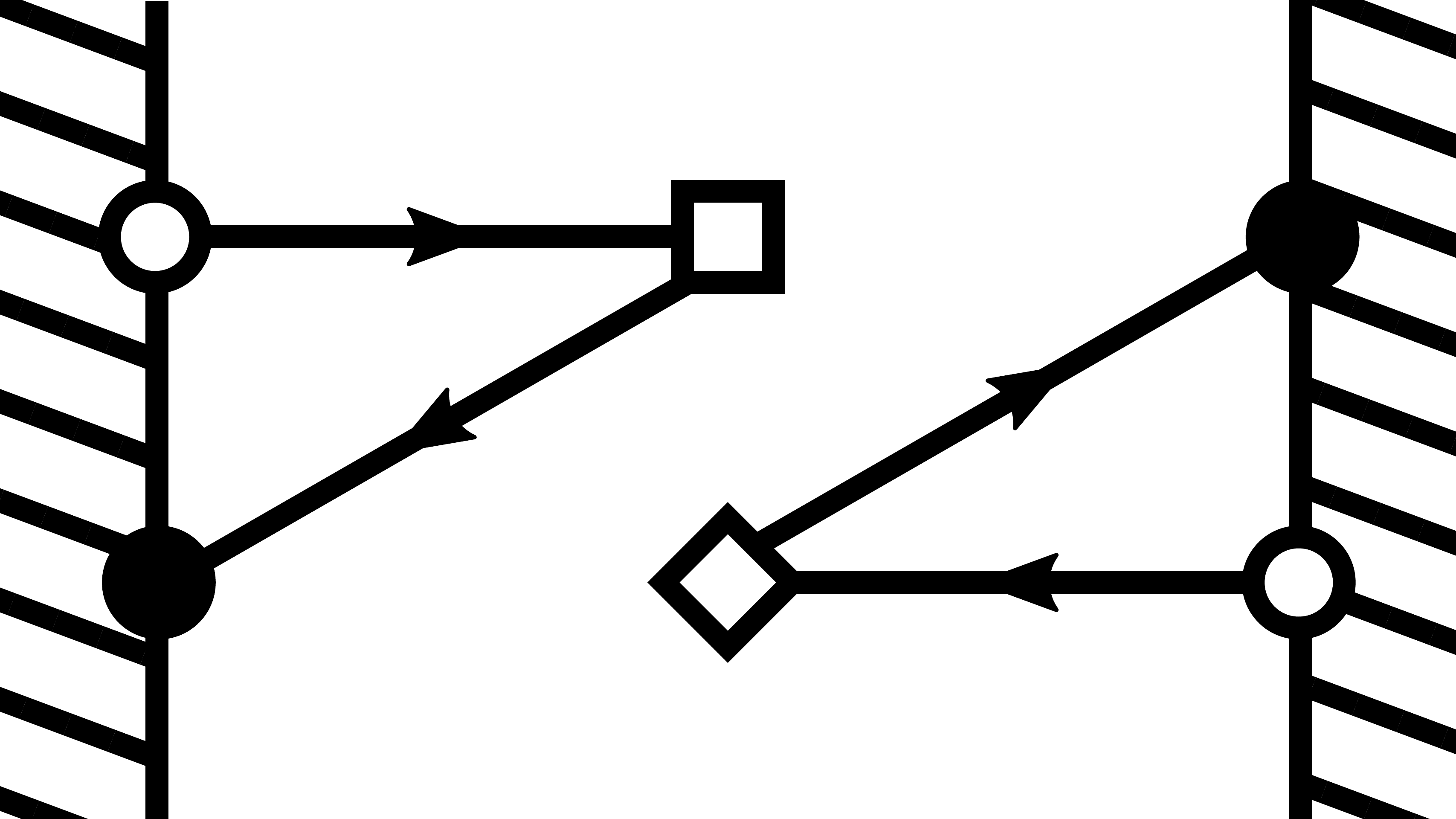}
		\caption{Diagram $F_{\mathrm{AP};7}$}
	\end{subfigure}
	\begin{subfigure}[b]{0.32\textwidth}
		\centering
		\includegraphics[width=0.85\textwidth]{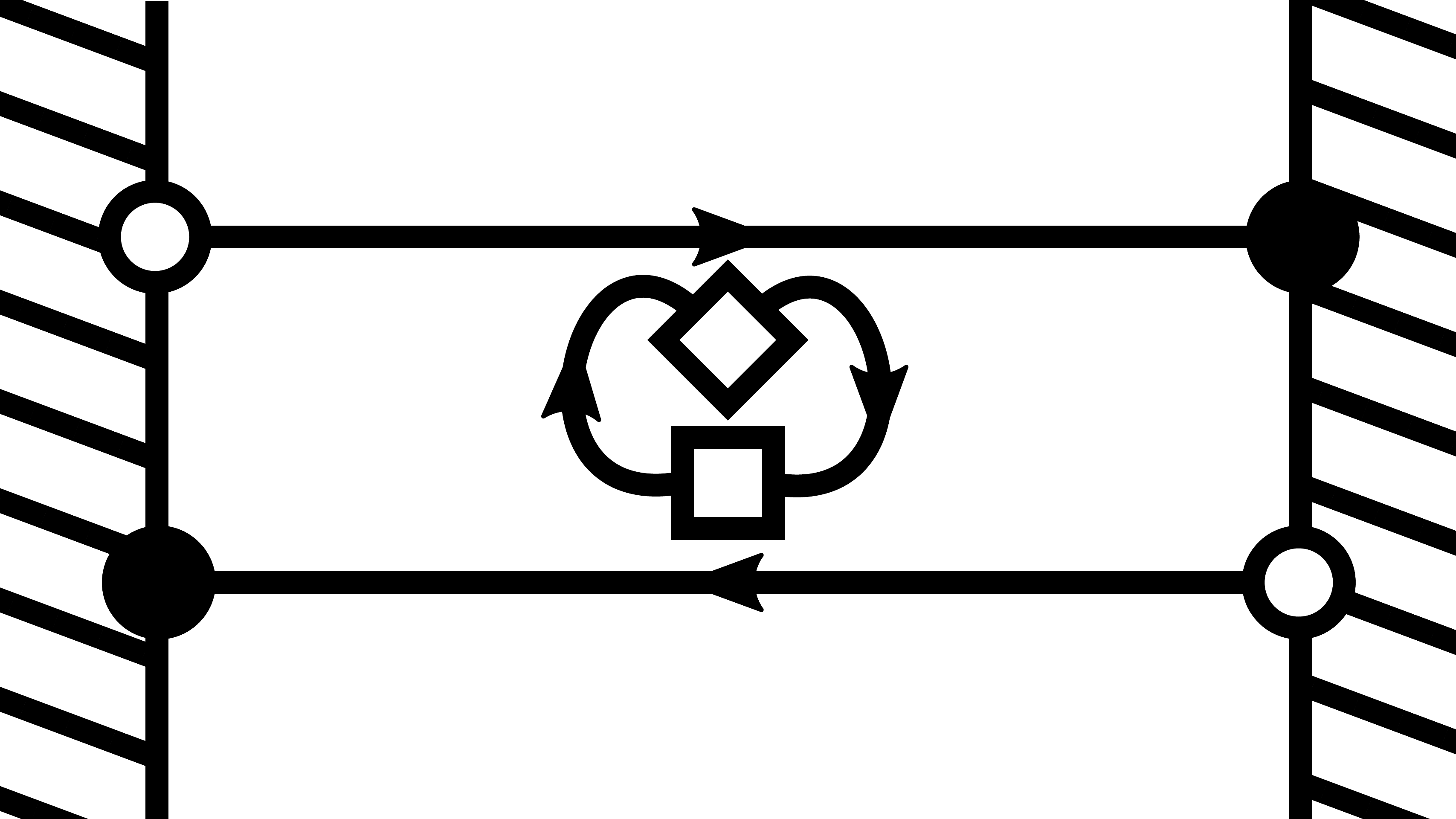}
		\caption{Diagram $F_{\mathrm{AP};9}$}
	\end{subfigure}
	\caption{The trace diagrams contributing to the expectation values of Table~\ref{tab:trBOOB}. Conventions are similar to those of Fig.~\ref{fig:WI-S}. The diamonds in the bulk represent the insertions of a pseudoscalar operator $P(y)$. The squares in the bulk represent the insertions of an axial current $A_0(x)$ or a pseudoscalar operator $P(x)$ (giving rise to the Dirac matrices $\gamma_0 \gamma_5$ or $\gamma_5$, respectively). Quark-connected diagrams $F_{\mathrm{AP};3}, F_{\mathrm{AP};5}, F_{\mathrm{AP};1}$ are single traces, formed by starting from any point and following the lines (quark propagators) around until we close the loop. Quark-disconnected diagrams $F_{\mathrm{AP};8}, F_{\mathrm{AP};7}, F_{\mathrm{AP};9}$ are products of two traces. Diagrams $F_{\mathrm{AP};2}, F_{\mathrm{AP};4}, F_{\mathrm{AP};6}$ are not shown, as they are related to $F_{\mathrm{AP};1}, F_{\mathrm{AP};3}, F_{\mathrm{AP};5}$; cf. Eqs.~(\ref{eq:F4T4}).
	}
	\label{fig:4point}
\end{figure*}
\begin{align}
\begin{split}
&a^6 \sum_{\mathbf{x}, \mathbf{y}} \langle \mathcal{O}^{\prime a} \, A_0^b(x) \, P^c(y) \,\mathcal{O}^d\rangle\\
=\,&a^{18}\sum_{k=1}^{9}T_k^{abcd} F_{\mathrm{AP};k}(x_0,y_0) \,.
\label{eq:TkFAPk}
\end{split}
\end{align}
Upon performing the Wick contractions, each correlation function is expressed as the sum of $9$ terms. They are products of traces of flavour matrices (denoted as $T_k^{abcd}$) and traces of loops of quark propagators averaged over gauge field configurations (denoted as $F_{\mathrm{AP};k}(x_0,y_0)$). The former traces are defined as:

{\small
\begin{align}
T_1^{abcd} &\equiv \Tr(T^aT^bT^cT^d) \,, \quad T_2^{abcd} \equiv \Tr(T^aT^dT^cT^b) \,, \label{eq:T12} \\
T_3^{abcd} &\equiv \Tr(T^aT^bT^dT^c) \,, \quad T_4^{abcd} \equiv \Tr(T^aT^cT^dT^b) \,,\label{eq:T34} \\
T_5^{abcd} &\equiv \Tr(T^aT^cT^bT^d) \,, \quad T_6^{abcd} \equiv \Tr(T^aT^dT^bT^c) \,, \label{eq:T56} \\
T_7^{abcd} &\equiv \Tr(T^aT^b)\Tr(T^dT^c) \,,\label{eq:T7} \\
T_8^{abcd} &\equiv \Tr(T^aT^c)\Tr(T^dT^b) \,,\label{eq:T8} \\
T_9^{abcd} &\equiv \Tr(T^aT^d)\Tr(T^cT^b) \label{eq:T9} \,,
\end{align}
}while the latter ones are also given in Table~\ref{tab:trBOOB}.
\begin{table*}[h!]
\begin{align*}
F_{\mathrm{S;1}}(y_0) =& \phantom{-}\,\,\sum_{\bf y} \sum_{\bf u,v,u^\prime, v^\prime} \left\langle \tr \left\lbrace [\zeta^\prime({\bf v^\prime}) \bar \zeta({\bf u})]_{\rm F} \gamma_5 [\zeta({\bf v}) \bar \psi(y)]_{\rm F} [\psi(y)  \bar \zeta^\prime({\bf u^\prime})]_{\rm F} \gamma_5 \right\rbrace  \right\rangle \\
F_{\mathrm{S;2}}(y_0) =& \phantom{-}\,\,\sum_{\bf y} \sum_{\bf u,v,u^\prime, v^\prime}  \left\langle \tr  \left\lbrace [\zeta^\prime({\bf v^\prime}) \bar \psi(y)]_{\rm F} [\psi(y) \bar \zeta({\bf u})]_{\rm F} \gamma_5 [\zeta({\bf v}) \bar \zeta^\prime({\bf u^\prime})]_{\rm F} \gamma_5  \right\rbrace  \right\rangle \\
\\
F_{\mathrm{AP};1}(x_0,y_0)=&-\,\sum_{\mathbf{x}, \mathbf{y}} \sum_{\mathbf{u}, \mathbf{v},\mathbf{u}^\prime,\mathbf{v}^\prime}\left\langle \mathrm{tr}\left\lbrace[\zeta^\prime (\mathbf{v}^\prime)\bar{\psi}(x)]\gamma_0 \gamma_5 [\psi(x)\bar{\psi}(y)]\gamma_5[\psi(y)\bar{\zeta}(\mathbf{u})]\gamma_5[\zeta(\mathbf{v})\bar{\zeta}^\prime (\mathbf{u}^\prime)]\gamma_5 \right\rbrace  \right\rangle \\
F_{\mathrm{AP};2}(x_0,y_0)=&-\,\sum_{\mathbf{x}, \mathbf{y}} \sum_{\mathbf{u}, \mathbf{v},\mathbf{u}^\prime,\mathbf{v}^\prime}\left\langle \mathrm{tr}\left\lbrace[\zeta^\prime (\mathbf{v}^\prime)\bar{\zeta}(\mathbf{u})]\gamma_5 [\zeta(\mathbf{v})\bar{\psi}(y)]\gamma_5[\psi(y)\bar{\psi}(x)]\gamma_0 \gamma_5[\psi(x)\bar{\zeta}^\prime (\mathbf{u}^\prime)]\gamma_5 \right\rbrace  \right\rangle  \\
F_{\mathrm{AP};3}(x_0,y_0)=&-\,\sum_{\mathbf{x}, \mathbf{y}} \sum_{\mathbf{u}, \mathbf{v},\mathbf{u}^\prime,\mathbf{v}^\prime}\left\langle \mathrm{tr}\left\lbrace[\zeta^\prime (\mathbf{v}^\prime)\bar{\psi}(x)]\gamma_0 \gamma_5 [\psi(x)\bar{\zeta}(\mathbf{u})]\gamma_5[{\zeta}(\mathbf{v})\bar{\psi}(y)]\gamma_5[\psi(y)\bar{\zeta}^\prime (\mathbf{u}^\prime)]\gamma_5 \right\rbrace  \right\rangle  \\
F_{\mathrm{AP};4}(x_0,y_0)=&-\,\sum_{\mathbf{x}, \mathbf{y}} \sum_{\mathbf{u}, \mathbf{v},\mathbf{u}^\prime,\mathbf{v}^\prime}\left\langle \mathrm{tr}\left\lbrace[\zeta^\prime (\mathbf{v}^\prime)\bar{\psi}(y)]\gamma_5 [\psi(y)\bar{\zeta}(\mathbf{u})]\gamma_5[{\zeta}(\mathbf{v})\bar{\psi}(x)]\gamma_0 \gamma_5[\psi(x)\bar{\zeta}^\prime (\mathbf{u}^\prime)]\gamma_5 \right\rbrace  \right\rangle \\
F_{\mathrm{AP};5}(x_0,y_0)=&-\,\sum_{\mathbf{x}, \mathbf{y}} \sum_{\mathbf{u}, \mathbf{v},\mathbf{u}^\prime,\mathbf{v}^\prime}\left\langle \mathrm{tr}\left\lbrace[\zeta^\prime (\mathbf{v}^\prime)\bar{\psi}(y)]\gamma_5 [\psi(y)\bar{\psi}(x)]\gamma_0 \gamma_5[\psi(x)\bar{\zeta}(\mathbf{u})]\gamma_5[\zeta(\mathbf{v})\bar{\zeta}^\prime (\mathbf{u}^\prime)]\gamma_5 \right\rbrace  \right\rangle  \\
F_{\mathrm{AP};6}(x_0,y_0)=&-\,\sum_{\mathbf{x}, \mathbf{y}} \sum_{\mathbf{u}, \mathbf{v},\mathbf{u}^\prime,\mathbf{v}^\prime}\left\langle \mathrm{tr}\left\lbrace[\zeta^\prime (\mathbf{v}^\prime)\bar{\zeta}(\mathbf{u})]\gamma_5 [\zeta(\mathbf{v})\bar{\psi}(x)]\gamma_0 \gamma_5[\psi(x)\bar{\psi}(y)]\gamma_5[\psi(y)\bar{\zeta}^\prime (\mathbf{u}^\prime)]\gamma_5 \right\rbrace  \right\rangle  \\
F_{\mathrm{AP};7}(x_0,y_0)=&+\sum_{\mathbf{x}, \mathbf{y}} \sum_{\mathbf{u}, \mathbf{v},\mathbf{u}^\prime,\mathbf{v}^\prime}\left\langle \mathrm{tr}\left\lbrace[\zeta^\prime (\mathbf{v}^\prime)\bar{\psi}(x)]\gamma_0 \gamma_5 [\psi(x)\bar{\zeta}^\prime (\mathbf{u}^\prime)]\gamma_5\right\rbrace \mathrm{tr}\left\lbrace [\psi(y)\bar{\zeta}(\mathbf{u})]\gamma_5 [\zeta(\mathbf{v})\bar{\psi}(y)]\gamma_5 \right\rbrace  \right\rangle  \\
F_{\mathrm{AP};8}(x_0,y_0)=&+\sum_{\mathbf{x}, \mathbf{y}} \sum_{\mathbf{u}, \mathbf{v},\mathbf{u}^\prime,\mathbf{v}^\prime}\left\langle \mathrm{tr}\left\lbrace[\zeta^\prime (\mathbf{v}^\prime)\bar{\psi}(y)]\gamma_5 [\psi(y)\bar{\zeta}^\prime (\mathbf{u}^\prime)]\gamma_5\right\rbrace \mathrm{tr}\left\lbrace [\psi(x)\bar{\zeta}(\mathbf{u})]\gamma_5 [\zeta(\mathbf{v})\bar{\psi}(x)]\gamma_0 \gamma_5 \right\rbrace  \right\rangle \\
F_{\mathrm{AP};9}(x_0,y_0)=&+\sum_{\mathbf{x}, \mathbf{y}} \sum_{\mathbf{u}, \mathbf{v},\mathbf{u}^\prime,\mathbf{v}^\prime}\left\langle \mathrm{tr}\left\lbrace[\zeta^\prime (\mathbf{v}^\prime)\bar{\zeta}(\mathbf{u})]\gamma_5 [\zeta(\mathbf{v})\bar{\zeta}^\prime (\mathbf{u}^\prime)]\gamma_5\right\rbrace \mathrm{tr}\left\lbrace [\psi(x)\bar{\psi}(y)]\gamma_5 [\psi(y)\bar{\psi}(x)]\gamma_0 \gamma_5 \right\rbrace  \right\rangle
\end{align*}
\caption{Mathematical expressions for the diagrams $ F_{\mathrm{S;}k}$ depicted in Fig.~\ref{fig:WI-S} and the diagrams $F_{\mathrm{AP};k}$ depicted in Fig.~\ref{fig:4point}.}
\label{tab:trBOOB}
\end{table*}
The spin-colour trace diagrams are shown in Fig.~\ref{fig:4point}. We see that there are six quark-connceted diagrams, and three quark-disconnected ones.  The condition $b \neq c$ implies that $T_9F_{\mathrm{AP};9}(x_0,y_0) =0$, due to the vanishing of $\mathrm{Tr}(T^cT^b)$. From Eq. (\ref{eq:TTnorm}) we see that $T_k^{abcd}$ for $k=7,8$ are real.

Once more we combine $\gamma_5$-Hermiticity, charge conjugation invariance, and Eq.~(\ref{eq:trTTTT}), to obtain for the l.h.s.\ of the Ward identity~(\ref{eq:SPImassintlatt}):

{
\small
\begin{align}
&\mathrm{WI~l.h.s.} = Z_\mathrm{A} Z_\mathrm{P} \,\, a^{18} \times \nonumber\\
& \Bigg [  \sum_{k=1,3,5} 2  \Re(T_k^{abcd}) \Big \{  F_{\mathrm{AP};k}(y_0+t,y_0) - F_{\mathrm{AP};k}(y_0-t,y_0) \Big \} \nonumber\\
& +  \sum_{k=7}^8  T_k^{abcd} \Big \{ F_{\mathrm{AP};k}(y_0+t,y_0) - F_{\mathrm{AP};k}(y_0-t,y_0) \Big \} \Bigg ] \,.
\label{eq:WIlhs}
\end{align}
}Note that correlation functions $F_{\mathrm{AP};k}$ are real for $k=1, \ldots, 9$. See \ref{app:corr-funct-symm} for more details. We will use a somewhat more compact notation, defining 
\begin{equation}
\label{eq:Deltadef}
\Delta_k(y_0,t)  \equiv  F_{\mathrm{AP};k}(y_0+t,y_0) - F_{\mathrm{AP};k}(y_0-t,y_0) \,.
\end{equation}
Collecting Eqs.~(\ref{eq:WIrhs}), (\ref{eq:WIlhs}), and (\ref{eq:Deltadef}), we write the Ward identity~(\ref{eq:SPImassintlatt}) in the chiral limit as:
\begin{align}
 &a^3 Z_\mathrm{A} Z_\mathrm{P} \times \nonumber\\
& \Bigg [ \sum_{k=1,3,5} 2  \Re(T_k^{abcd}) \Delta_k(y_0,t)
+ \sum_{k=7,8} T_k^{abcd}  \Delta_k(y_0,t) \Bigg ]\nonumber\\
=&-\dfrac{Z_{\mathrm S}}{2} d^{bce} d^{ade} \Re\Big [ F_{\mathrm{S;1}}(y_0) \Big ] + \rmO(a^2) \,.
\label{eq:WImaster-Delta}
\end{align}

In order to keep the equation simple, we have not shown the mass-dependent terms with two pseudoscalar density insertions, appearing in Eq.~(\ref{eq:SPImassintlatt}). These terms are included in the numerical analysis, which is carried out close to, but not strictly at the chiral limit. The reader should have no difficulty convincing himself that they are exactly analogous to $F_{\mathrm{AP};k}(y_0+t,y_0)$ and $ F_{\mathrm{AP};k}(y_0-t,y_0)$ appearing above. Their net effect is to add extra mass-dependent contributions to the $\Delta_k(y_0,t)$ functions. From now on, the $\Delta_k(y_0,t)$ functions are meant to include these contributions, proportional to the quark mass. Consequently, the uncertainty on the r.h.s. of Eq.~(\ref{eq:WImaster-Delta}) becomes $\rmO(am,a^2)$.

It is interesting to compare the Ward identities we have derived here to the one introduced in Ref.~\cite{Luscher:1996jn} for the determination of $Z_\mathrm{A}$. The former are valid for $\NF \geq 3$, while the latter for $\Nf \geq 2$. The Ward identity of Ref.~\cite{Luscher:1996jn} involves correlation functions with two axial current insertions in the bulk. In our case we have more complicated contributions, consisting of time-differences of correlation functions with one axial current and one pseudoscalar density insertion.

\section{Determination of  \texorpdfstring{$Z_\mathrm{S}/(Z_\mathrm{P}Z_\mathrm{A})$}{ZS/(ZP ZA)} from Ward identities}
\label{sec:corr-funct}

Ward identity~(\ref{eq:WImaster-Delta}) is a master equation, from which a plethora of relations arise for specific choices of flavour indices $a,b,c,d$.
In what follows, each of them will be distinguished by the label WI($abcd$). Not all of them are suitable for the determination of $Z_\mathrm{S}/Z_\mathrm{P}$. 
The following constraints need to be imposed:
\begin{enumerate}[i]
\item [(i)] \label{it:bneqc}
$b \neq c$; this ensures the suppression of the scalar term in Eq.~(\ref{eq:deltaP});
\item [(ii)] \label{it:dbceneq0}
$d^{bce} \neq 0$ and $d^{ade} \neq 0$, so that the r.h.s.\ of Eq.~(\ref{eq:WImaster-Delta}) does not vanish. Note that once $b,c$ are fixed, property A in \ref{app:sun} ensures that $d^{bce} \neq 0$ for a single value of $e$. Thus the summation over $e$ on the r.h.s.\ of our master equation is trivial and the requirement $d^{bce} d^{ade} \neq 0$ is satisfied for at most a single value of $e$;
\item [(iii)] \label{it:fbceneq0}
$f^{bce} = 0$ for the choice of indices $b,c,e$ for which $d^{bce} \neq 0$; $f^{ade} = 0$ for the choice of indices $a,d,e$ for which $d^{ade} \neq 0$.
This follows from property B in \ref{app:sun}.
\end{enumerate}

In spite of these constraints, a lot of freedom remains in the choice of flavour indices, resulting in many Ward identities. They are relations between the correlation functions of the master equation, which can be solved for $Z_\mathrm{S}/(Z_\mathrm{P}Z_\mathrm{A})$. These Ward identities can be grouped into different equivalence classes. Each class consists of several identities WI($abcd$) with different flavour indices $a,\ldots,d$, but identical flavour factors $\Re(T_k)$ ($k=1,3,5,7,8$), and thus the same Eq.~(\ref{eq:WImaster-Delta}). Therefore, the same $Z_\mathrm{S}/(Z_\mathrm{P}Z_\mathrm{A})$ estimate is obtained from all Ward identities of the same equivalence class. Estimates of  $Z_\mathrm{S}/(Z_\mathrm{P}Z_\mathrm{A})$ from Ward identities of different classes differ by discretisation effects. 

The combinations of conditions (i)--(iii) simmer down to the choice of flavour indices $(a,b,c,d)$, with $b \neq c$, such that $d^{bce} d^{ade}\neq 0$. We systematically investigated the choices of flavour indices which fulfill these conditions with a computer algebra program and grouped them into the equivalence classes which are tabulated in Table~\ref{tab:WIclasses}. These results depend on the $su(\NF)$ Gell-Mann matrix definitions of \ref{app:sun}. Some interesting observations are:
\begin{itemize}
\item There are pairs of equivalence classes that have the same number of elements. Examples are WI(1245) paired to WI(1425), WI(1144) paired to WI(1414) etc. These pairs of classes are separated by a single horizontal line in Table~\ref{tab:WIclasses}. Class WI(1468) does not have a partner.
\item The flavour factors $\Re(T_k)$ for $(k = 1,3,5)$, $T_7$, and $T_8$ of paired classes
have closely related numerical values; see Table~\ref{tab:F-classes}. 
We will see below how this leads to useful relations between certain $\Delta_k$ functions.
\item The quark disconnected traces $\Delta_7$ and $\Delta_8$ do not contribute to the equivalence classes of the top half of Table~\ref{tab:WIclasses} (separated by a triple line from the bottom half).
\end{itemize}

\begin{table*}
	\centering
	\begin{tabular*}{\textwidth}{@{\extracolsep{\fill}}cl@{}}
		\toprule
		\begin{tabular}{c}
		Equivalence	\\
		class label
		\end{tabular}
		   & \multicolumn{1}{c}{Equivalence class elements}\\
		\midrule
		 1245    & 1245 1254 1267 1276 1346 1357 1364 1375 2145 2154 2167 2176 2347 2356 2365 2374 3146 3157 3164 3175 \\
		 & 3247 3256 3265 3274 4512 4521 4567 4576 4613 4631 4723 4732 5412 5421 5467 5476 5623 5632 5713 5731 \\
		 & 6413 6431 6523 6532 6712 6721 6745 6754 7423 7432 7513 7531 7612 7621 7645 7654 \\
		\midrule
		 1425    & 1425 1436 1524 1537 1627 1634 1726 1735 2415 2437 2514 2536 2617 2635 2716 2734 3416 3427 3517 3526 \\
		 & 3614 3625 3715 3724 4152 4163 4251 4273 4361 4372 4657 4756 5142 5173 5241 5263 5362 5371 5647 5746 \\
		 & 6143 6172 6253 6271 6341 6352 6475 6574 7153 7162 7243 7261 7342 7351 7465 7564                     \\
		\midrule
		\midrule
		 1486    & 1486 1587 1684 1785 2487 2586 2685 2784 3484 3585 3686 3787 4168 4278 4348 4843 4861 4872 5178 5268 \\
		 & 5358 5853 5862 5871 6148 6258 6368 6841 6852 6863 7158 7248 7378 7842 7851 7873 8416 8427 8434 8517 \\
		 & 8526 8535 8614 8625 8636 8715 8724 8737 \\
		\midrule
		 1846    & 1846 1857 1864 1875 2847 2856 2865 2874 3844 3855 3866 3877 4438 4483 4618 4681 4728 4782 5538 5583 \\
		 & 5628 5682 5718 5781 6418 6481 6528 6582 6638 6683 7428 7482 7518 7581 7738 7783 8146 8157 8164 8175 \\
		 & 8247 8256 8265 8274 8344 8355 8366 8377 \\
		\midrule
		\midrule
		 1468    & 1468 1578 1648 1758 2478 2568 2658 2748 4186 4287 4384 4816 4827 4834 5187 5286 5385 5817 5826 5835 \\
		 & 6184 6285 6386 6814 6825 6836 7185 7284 7387 7815 7824 7837 8461 8472 8562 8571 8641 8652 8742 8751 \\
		\midrule
		\midrule
		\midrule
		 1144    & 1144 1155 1166 1177 2244 2255 2266 2277 3344 3355 3366 3377 4411 4422 4433 4466 4477 5511 5522 5533 \\
		 & 5566 5577 6611 6622 6633 6644 6655 7711 7722 7733 7744 7755 \\
		\midrule
		 1414    & 1414 1515 1616 1717 2424 2525 2626 2727 3434 3535 3636 3737 4141 4242 4343 4646 4747 5151 5252 5353 \\
		 & 5656 5757 6161 6262 6363 6464 6565 7171 7272 7373 7474 7575 \\
		\midrule
		\midrule
		1188    & 1188 2288 3388 8811 8822 8833 \\
		\midrule
		1818    & 1818 2828 3838 8181 8282 8383 \\
		\midrule
		\midrule
		4488    & 4488 5588 6688 7788 8844 8855 8866 8877 \\
		\midrule
		4848    & 4848 5858 6868 7878 8484 8585 8686 8787 \\                        
		\bottomrule
	\end{tabular*}
    \caption{Ward identities WI($abcd$) grouped into equivalence classes. Each class is labeled by four flavour indices $abcd$, of a representative element,
    	listed in the leftmost column. All elements of the same class are grouped to the right. For more explanations, see text.}
    \label{tab:WIclasses}
\end{table*}

In Table~\ref{tab:F-classes} we collect the flavour factors $\Re(T_k)$ ($k=1,3,5$), $T_7$, and $T_8$ for each class. Depending on the choice of flavour indices $a$,$b$,$c$,$d$, some of these flavour factors vanish. This simplifies the resulting Ward identity. Also here the top part of the Table (separated by a double line from the bottom half) lists the Ward identities without $\Delta_7$- and $\Delta_8$-type contributions.
\begin{table*}    
\centering
\renewcommand{\arraystretch}{1.25}
\setlength{\tabcolsep}{3pt}
    \begin{tabular*}{\textwidth}{@{\extracolsep{\fill}}ccccccc@{}}
         \toprule
         WI(abcd) & $\Re(T_1^{abcd})$ & $\Re(T_3^{abcd})$ & $\Re(T_5^{abcd})$ & $T_7^{abcd}$ & $T_8^{abcd}$& $d^{bce}d^{ade}$  \\
         \midrule
         WI(1245) & $-1/16$ & $1/16$ & $0$ & $0$ & $0$ & $-1/4$ \\
         WI(1425) & $0$ & $1/16$ & $-1/16$ & $0$ & $0$ & $-1/4$ \\
         \midrule
         WI(1486) & $-\sqrt{3}/24$ & $\sqrt{3}/48$ & $\sqrt{3}/48$ & 0 & 0 & $-\sqrt{3}/12$ \\
         WI(1846) & $\sqrt{3}/48$ & $\sqrt{3}/48$ & $-\sqrt{3}/24$  & $0$ & $0$ & $-\sqrt{3}/12$ \\
         \midrule
         WI(1468) & $\sqrt{3}/48$ & $-\sqrt{3}/24$ & $\sqrt{3}/48$ & $0$ & $0$ & $\sqrt{3}/6$ \\
		 \midrule \midrule
         WI(1144) & $1/16$ & $1/16$ & $0$ & $1/4$ & $0$ & $1/4$ \\
         WI(1414) & $0$ & $1/16$ & $1/16$ & $0$ & $1/4$ & $1/4$ \\
         \midrule
         WI(1188) & $1/24$ & $1/24$ & $1/24$ & $1/4$ & $0$ & $1/3$ \\
         WI(1818) & $1/24$ & $1/24$ & $1/24$ & $0$ & $1/4$ & $1/3$ \\
         \midrule
         WI(4488) & $5/48$ & $5/48$ & $-1/12$ & $1/4$ & $0$ & $1/12$ \\
         WI(4848) & $-1/12$ & $5/48$ & $5/48$ & $0$ & $1/4$ & $1/12$ \\
		\bottomrule
    \end{tabular*}
    \caption{Classes of Ward identities (first column), the corresponding flavour factors of Eq.~(\ref{eq:WImaster-Delta}) (columns 2 to 6) and the product of symmetric
    tensors $d$ of the same equation (last column).}
    \label{tab:F-classes}
\end{table*}

There are two possible ways of using the 11 Ward identities of Table~\ref{tab:F-classes}. A first approach would be to determine $Z_\mathrm{S}/(Z_\mathrm{P}Z_\mathrm{A})$ from each of the 11 variants of Eq.~(\ref{eq:WImaster-Delta}). In principle these determinations differ by $\rmO(am,a^2)$ effects and that should provide a handle for a good control of the related systematics. However, in practice the different $Z_\mathrm{S}/(Z_\mathrm{P}Z_\mathrm{A})$ results are all obtained from the same configuration ensembles and are thus strongly correlated. Moreover, paired Ward identities (in the sense discussed above; cf. Table~\ref{tab:WIclasses}) have very similar relations between their $\Delta_k$-terms and this also leads to very similar $Z$-ratios.

A second approach would be to combine these Ward identities in order to first obtain relations between the various $\Delta_k$-terms. These would be true up to $\rmO(am,a^2)$ at fixed gauge coupling, and once established, would simplify the equation(s) relating $Z_\mathrm{S}/(Z_\mathrm{P}Z_\mathrm{A})$ to the $\Delta_k$'s.
In this spirit we proceed as follows:\\
(i) Starting from Ward identities without quark disconnected contributions (i.e., with $\Re(T_7) = \Re(T_8)=0$; top part of Table~\ref{tab:F-classes}), we combine the pair WI(1245) and WI(1425) to obtain:
\begin{align}
& \Delta_1(y_0,t) =  \Delta_5(y_0,t) +\rmO(am,a^2) \,,
\label{eq:D1=D5}
\end{align}
\begin{align}
\begin{split}
& Z_\mathrm{A} Z_\mathrm{P} a^3 \big [ \Delta_1(y_0,t) -  \Delta_3(y_0,t) \big ] \\
= &- Z_\mathrm{S} \Re \big [ F_{\mathrm{S;1}}(y_0) \big ] + \rmO(am,a^2) \,.
\label{eq:D1-D3=rhs}
\end{split}
\end{align}
Note that by combining the pair WI(1486) and WI(1846) we also obtain the above expressions, so this pair does not provide extra information.\\
(ii) WI(1468), which has no partner, is written, in terms of the $\Delta$'s defined in Eq.~(\ref{eq:Deltadef}), as:
\begin{align}
\begin{split}
&Z_\mathrm{A} Z_\mathrm{P} a^3 \big [ \Delta_1(y_0,t) -  2 \Delta_3(y_0,t) + \Delta_5(y_0,t) \big ] \\
=& -2 \, Z_\mathrm{S} \Re \big [ F_{\mathrm{S;1}}(y_0) \big ] + \rmO(am,a^2) \,.
\label{eq:D1-2D3+D5=rhs}
\end{split}
\end{align}
This on its own determines the ratio $Z_\mathrm{S}/(Z_\mathrm{P}Z_\mathrm{A})$. Note that combined with Eq.~(\ref{eq:D1=D5}), it gives us Eq.~(\ref{eq:D1-D3=rhs}).
Our conclusion is that all Ward identities with $\Re(T_7) = \Re(T_8)=0$ reduce to the equality $\Delta_1 =  \Delta_5$ (i.e., diagrams $F_{\mathrm{AP};1}$ and $F_{\mathrm{AP};5}$ of Fig.~\ref{fig:4point} are related) and a single Ward identity, from which $Z_\mathrm{S}/(Z_\mathrm{P}Z_\mathrm{A})$ may be computed.
\\
(iii) Passing to Ward identities with quark-disconnected contributions (bottom part of Table~\ref{tab:F-classes}), we combine the pair WI(1188) and WI(1818) to obtain:
\begin{align}
& \Delta_7(y_0,t)  =  \Delta_8(y_0,t) +\rmO(am,a^2) \,,
\label{eq:D7=D8}
\end{align}
\begin{align}
\begin{split}
& Z_\mathrm{A} Z_\mathrm{P} a^3 \big [ 2 \Delta_1(y_0,t) +  \Delta_3(y_0,t) +  3 \Delta_7(y_0,t) \big ] \\
=& -2 Z_\mathrm{S} \Re \big [ F_{\mathrm{S;1}}(y_0) \big ] + \rmO(am,a^2) \,,
\label{eq:2D1+D3+3D7=rhs}
\end{split}
\end{align}
where Eq.~(\ref{eq:D1=D5}) has also been used to arrive at Eq.~(\ref{eq:2D1+D3+3D7=rhs}).\\
(iv) Similarly, the pair WI(1144) and WI(1414) combine to give
\begin{align}
\begin{split}
& \Delta_1(y_0,t) + 2 \Delta_7(y_0,t) \\
=\,&  \Delta_5(y_0,t) + 2 \Delta_8(y_0,t) + \rmO(am,a^2) \,,
\label{eq:D1+2D7=D5+2D8}
\end{split}
\end{align}
\begin{align}
\begin{split}
& Z_\mathrm{A} Z_\mathrm{P} a^3 2 \big [ \Delta_1(y_0,t) + \Delta_3(y_0,t) +  2 \Delta_7(y_0,t) \big ] \\
=& -2 Z_\mathrm{S} \Re \big [ F_{\mathrm{S;1}}(y_0) \big ] + \rmO(am,a^2) \,.
\label{eq:2D1+2D3+4D7=rhs}
\end{split}
\end{align}
Eq.~(\ref{eq:D1+2D7=D5+2D8}) carries no new information, as it is a combination of Eqs.~(\ref{eq:D1=D5}) and (\ref{eq:D7=D8}).\\
(v) If we now combine Eqs.~(\ref{eq:2D1+D3+3D7=rhs}) and (\ref{eq:2D1+2D3+4D7=rhs}), we obtain again Eq.~(\ref{eq:D1-D3=rhs}) and the new relation
\begin{equation}
\Delta_3(y_0,t)  =  - \Delta_7(y_0,t) +\rmO(am,a^2) \,.
\label{eq:D3=-D7} \\
\end{equation}
The bottom line is that, up to $\rmO(am,a^2)$ discretisation effects, the 11 Ward identities corresponding to the entries of Table~\ref{tab:F-classes} are not all independent. They can be combined to give three relations between the functions $\Delta_k$, which depend on traces of valence quark propagators, without references to flavour traces; these are Eqs.~(\ref{eq:D1=D5}), (\ref{eq:D7=D8}), and (\ref{eq:D3=-D7})\footnote{As an aside we note that Eqs.~(\ref{eq:D1=D5}) and (\ref{eq:D7=D8}) relate correlation functions of similar topology (quark-connected or quark-disconnected ones). On the contrary, Eq.~(\ref{eq:D3=-D7}) is more intriguing, as it relates quark-connected to quark-disconnected diagrams.}. The extent to which these relations are fulfilled at non-zero lattice spacing is an indicator of the size of discretisation effects. Moreover, if we take them at face value, the remaining Ward identities~(\ref{eq:D1-D3=rhs}), (\ref{eq:D1-2D3+D5=rhs}), 
(\ref{eq:2D1+D3+3D7=rhs}), and (\ref{eq:2D1+2D3+4D7=rhs}) reduce to a single expression. Any of them can be used to provide estimates of the ratio $Z_\mathrm{S}/(Z_\mathrm{P}Z_\mathrm{A})$. We expect Eqs.~(\ref{eq:2D1+D3+3D7=rhs}), and (\ref{eq:2D1+2D3+4D7=rhs}) to be noisier, as they involve quark-disconnected diagrams. Eq.~(\ref{eq:D1-D3=rhs}) seems promising, as it only involves $\Delta_1$ and $\Delta_3$, but it cannot be excluded {\it a priori} that Eq.~(\ref{eq:2D1+D3+3D7=rhs}) turns out to be better behaved. This can only be decided by numerical investigation.

Of course, these considerations do not exhaust all possibilities. Any linear combination of the Ward identities considered above, possibly combined with the relations~(\ref{eq:D1=D5}), (\ref{eq:D7=D8}), (\ref{eq:D3=-D7}), can be used for the computation of $Z_\mathrm{S}/(Z_\mathrm{P}Z_\mathrm{A})$. For example, the linear combination ${\rm L}_1\equiv$~[WI($1245$)$-$WI($1425$)], combined with Eq.~(\ref{eq:D1=D5}) gives:
\begin{align}
\begin{split}
&Z_\mathrm{A} Z_\mathrm{P} a^3 \big [ \Delta_1(y_0,t) \big ] \\
=& - Z_\mathrm{S} \Re \big [ F_{\mathrm{S;1}}(y_0) \big ] + \rmO(am,a^2)\,.
\label{eq:L1}
\end{split}
\end{align}
The determination of $Z_\mathrm{S}/(Z_\mathrm{P}Z_\mathrm{A})$ from the above depends only on quark-connected diagrams. Similarly, the linear combination ${\rm L}_2 \equiv$~[$12$WI($1818$)$-8$WI($1414$)] gives:
\begin{align}
\begin{split}
&Z_\mathrm{A} Z_\mathrm{P} a^3 \big [ \Delta_1(y_0,t) + \Delta_8(y_0,t) \big ] \\
=& - Z_\mathrm{S} \Re \big [ F_{\mathrm{S;1}}(y_0) \big ] + \rmO(am,a^2)\,,
\label{eq:L2}
\end{split}
\end{align}
which yields a $Z_\mathrm{S}/(Z_\mathrm{P}Z_\mathrm{A})$ estimate from quark-connected and quark-disconnected diagrams. The last two expressions will be used in the following for numerical crosschecks.

\section{Numerical setup and results}
\label{section:results}
We investigate the proposed Ward identities on lattices with tree-level Symanzik improved gluons and Wilson-Clover quarks. The action coincides with the one used by CLS \cite{Bruno:2014jqa,Bali:2016umi,Mohler:2017wnb}. We employ Schrödinger functional boundary conditions in time, which enable us to simulate at quark masses close to the chiral point and control systematic effects related to the massless renormalisation framework. The details of this aspect are discussed in Subsection~\ref{sec:chiral_extrapolations}. 
Similar to the procedure in \cite{Bulava:2016ktf}, we construct boundary-to-boundary three- and four-point functions with pseudoscalar Schrödinger functional wall sources and use wavefunctions at the boundaries as explained in \cite{Bulava:2015bxa}. The statistical error analysis is performed using a python implementation of the $\Gamma$-method \cite{Wolff:2003sm} (exploiting information from the autocorrelation function) with automatic differentiation \cite{Ramos:2018vgu}.
\begin{table*}[tb]
\centering
\begin{tabular*}{\textwidth}{@{\extracolsep{\fill}}llllrll@{}}
	\toprule
 ID       & $L^3\times T/a^4$   &   $\beta$ &   $\kappa$ &   MDU & $\tau_\mathrm{exp}$   & $a$ in fm   \\
 \midrule
 A1k1     & $12^3\times 17$     &     3.3   &   0.13652  & 20480 & \phantom{0}1.031(71)             & 0.1045(18)  \\
 A1k3     & $12^3\times 17$     &     3.3   &   0.13648  &  6876 & \phantom{0}2.06(14)              & 0.1045(18)  \\
 A1k4     & $12^3\times 17$     &     3.3   &   0.1365   & 96640 & \phantom{0}1.031(71)             & 0.1045(18)  \\
 \midrule
 E1k1     & $14^3\times 21$     &     3.414 &   0.1369   & 38400 & \phantom{0}1.61(12)              & 0.08381(68) \\
 E1k2     & $14^3\times 21$     &     3.414 &   0.13695  & 57600 & \phantom{0}1.61(12)              & 0.08381(68) \\
 \midrule
 B1k1     & $16^3\times 23$     &     3.512 &   0.137    & 20480 & \phantom{0}4.41(96)              & 0.06954(43) \\
 B1k2     & $16^3\times 23$     &     3.512 &   0.13703  &  8192 & \phantom{0}4.41(96)              & 0.06954(43) \\
 B1k3     & $16^3\times 23$     &     3.512 &   0.1371   & 16384 & \phantom{0}4.41(96)              & 0.06954(43) \\
 B1k4     & $16^3\times 23$     &     3.512 &   0.13714  & 27856 & \phantom{0}4.41(96)              & 0.06954(43) \\
 \midrule
 C1k1     & $20^3\times 29$     &     3.676 &   0.1368   &  7848 & 10.7(4.1)             & 0.05170(42) \\
 C1k2     & $20^3\times 29$     &     3.676 &   0.137    & 15232 & 10.7(4.1)             & 0.05170(42) \\
 C1k3     & $20^3\times 29$     &     3.676 &   0.13719  & 15472 & 10.7(4.1)             & 0.05170(42) \\
 \midrule
 D1k2     & $24^3\times 35$     &     3.81  &   0.13701  &  5360 & 62(14)                & 0.04175(70) \\
 D1k4     & $24^3\times 35$     &     3.81  &   0.137033 & 79664 & 31.0(7.0)             & 0.04175(70) \\
 \bottomrule
\end{tabular*}

\caption{
Summary of simulation parameters: the first column (ID) labels our gauge configuration ensembles,  the second column lists the lattice sizes $L^3 \times T/a^4$, the third one the inverse gauge couplings $\beta$, the fourth the Wilson hopping parameters $\kappa$, the fifth shows the total number of molecular dynamics units MDU, the sixth the autocorrelation time of the slowest mode $\tau_{\mathrm exp}$, and the last one the corresponding lattice spacing $a$, estimated from Ref.~\cite{Bruno:2016plf}.
}\label{tab:ens}
\end{table*}

The gauge ensembles used in this study are detailed in Table~\ref{tab:ens}. They coincide with the ones used in \cite{deDivitiis:2019xla} but for the ensemble C1k1. These are essentially the ensembles used in \cite{Bulava:2015bxa,Bulava:2016ktf} plus the ensembles A1k3, A1k4, B1k4, C1k1, D1k2 and D1k4, which were added to improve the chiral fits. For the two ensembles E1k1 and E1k2 the number of molecular dynamics units was increased by factor of more than $4$. The ensembles with volume $L^3\times T$ described above are designed to lie on a line of constant
physics (LCP), where the spatial extent of $L \approx 1.2\,\fm$ and
$T/L\approx 3/2$ are almost constant.
The Ward identity conditions which fix the ratio $Z_\mathrm{S}/(Z_\mathrm{P} Z_\mathrm{A})$ are imposed at constant physics, i.e., we require that all length scales in the correlation functions, which define a given condition formulated through one of the foregoing Ward identities, are kept fixed in physical units. Once this requirement is satisfied, only the lattice spacing $a$ changes as $g_0$ is varied. Consequently, renormalisation constants (as well as their ratios) extracted from different constant physics conditions are expected to rapidly approach an almost unique function of $g_0$ as $g_0\to 0$. For a more general discussion of the constant physics idea in a similar context see, e.g., Ref.~\cite{Fritzsch:2010aw}.

The initial tuning of this LCP was done based on the (universal)
2-loop beta-function as explained in Ref.~\cite{Bulava:2015bxa}. Thus the volume of the lattices varies by $\approx 10$\% over the range of couplings considered. However, using the results of Ref.~\cite{Bruno:2016plf}, we verified that this deviation is proportional to the lattice spacing $a$ and thus contributes to our quantity of interest only as a higher-order ambiguity\footnote{A more explicit quantitative investigation of violations of the constant
physical volume requirement by our Schrödinger functional ensembles,
demonstrating that it affects the Ward identity determination of improvement
coefficients and normalisation factors only beyond the order we are actually interested in, will be reported in \cite{Heitger2020}.}.

The simulations in this work suffer from critical slowing down of the
topological charge for smaller lattice spacings. This phenomenon,
often dubbed "topology freezing", could give unreliable results due to
an insufficient sampling of topological sectors.
We circumvent this problem by reweighting all data to the trivial
topological sector $Q = 0$ at the cost of decreasing the effective number of
configurations; see \cite{Bulava:2015bxa,Fritzsch:2013yxa} for a discussion. Furthermore we increase the statistical uncertainties by attaching a tail to the integrated autocorrelation functions as proposed in \cite{Schaefer:2010hu}. As measure for $\tau_\mathrm{exp}$, the autocorrelation time of the slowest mode in the simulation, we use the integrated autocorrelation time of the squared topological charge $Q^2$ extracted from the longest Monte Carlo chain for each value of $\beta$. The $\tau_\mathrm{exp}$-values for the individual ensembles can be found in Table~\ref{tab:ens}. 

In order to solve the Ward identity for $Z_\mathrm{S}/Z_\mathrm{P}$ we need non-perturbative knowledge of the non-singlet axial current renormalisation constant $Z_\mathrm{A}$ and the $\mathrm{O}(a)$ improvement coefficient $c_\mathrm{A}$. The constant $Z_\mathrm{A}$ was calculated on a subset of the gauge configurations in this work, Ref.~\cite{Bulava:2016ktf}, as well as in the chirally rotated Schrödinger functional, Ref.~\cite{DallaBrida:2018tpn}, which is a completely different determination. We prefer the results from the latter because of their smaller statistical uncertainties. The errors of $Z_\mathrm{A}$ are accounted for in quadrature when solving for $Z_\mathrm{S}/Z_\mathrm{P}$ in our Ward identity expressions. For $c_\mathrm{A}$ we use the results of \cite{Bulava:2015bxa}, without error, following standard practice.

In principle the ratio we would like to determine, as well as all correlation functions involved, depend on the $\mathrm{O}(a)$ improved coupling $\tilde{g}_0^2=g_0^2[1+ab_\mathrm{g}\mathrm{tr}\,M_\mathrm{q}/N_\mathrm{f}]$, where the coefficient $b_\mathrm{g}$ is only known at 1-loop perturbation theory \cite{Luscher:1996sc}. This issue is of no relevance here, as all normalisation conditions are imposed at zero quark mass. However, this should be kept in mind when using results obtained here in a different setting with non-vanishing sea quark masses.

In order to study the scaling behaviour of some of our results, we need the lattice spacings in physical units at the bare couplings used in this work. In Ref.~\cite{Bruno:2016plf}, such values are provided for couplings close to those in Table~\ref{tab:ens}; these enable us to extract the lattice spacings at our gauge couplings using a polynomial interpolation.

As additional cross checks we investigate the non-perturbative validity of the identities (\ref{eq:D1=D5}), (\ref{eq:D7=D8}) and (\ref{eq:D3=-D7}). The results can be found in \ref{app:np_identities}.

\subsection{Chiral extrapolation}
\label{sec:chiral_extrapolations}
From the plethora of possible renormalisation conditions listed in Section \ref{sec:corr-funct}, we single out a class labeled WI($1468$) to which only quark connected diagrams contribute and for which the statistical precision is best. We detail the analysis for this specific choice, but the same steps also apply to any other identity discussed in the following.
\begin{figure}[htb]
	\centering
	\includegraphics[width=\linewidth]{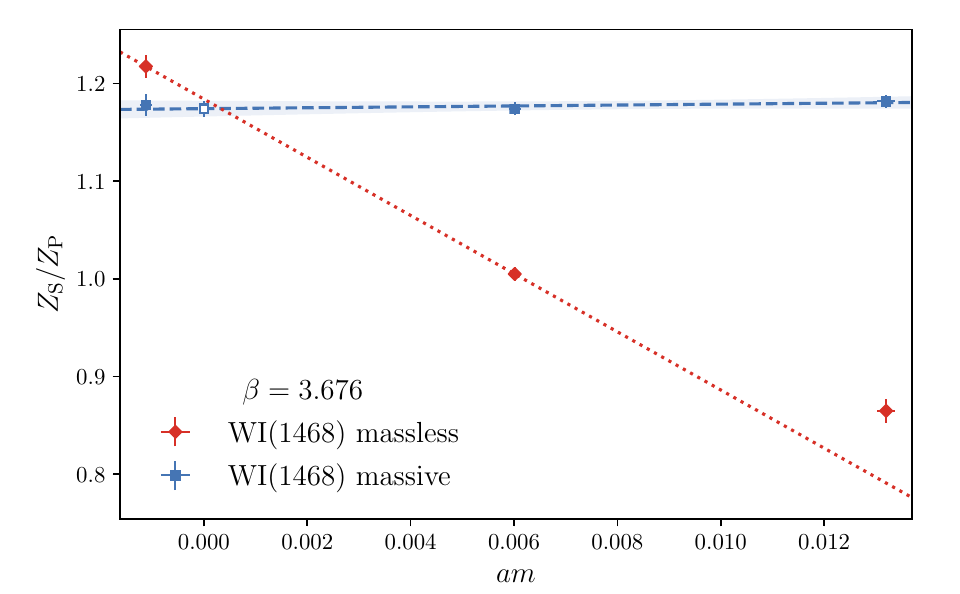}
	\caption{Comparison of the chiral extrapolation for WI(1468) at $\beta=3.676$ with and without the term proportional to the mass. In the massless case the data cannot be described by a linear function in $am$ for the full mass range. The dotted line visualises the chiral extrapolation of the massless data set excluding the outmost data point. When the mass term is included, the data shows no significant quark mass dependence. The slope of the linear fit function, shown as the dashed line, where the shaded area corresponds to the 1$\sigma$ uncertainty, is zero within error.}
	\label{fig:chiral_extrapolation}
\end{figure}
In order to obtain $Z_\mathrm{S}/Z_\mathrm{P}$ at vanishing quark mass, we extra- or interpolate the data at fixed bare coupling to the chiral point. For this procedure we employ the $\mathrm{O}(a)$ improved PCAC mass, which we average over the central third of the temporal extent of the lattice, similarly to what was done in Ref.~\cite{deDivitiis:2019xla}. This choice keeps the plateau length approximately constant in physical units. For the insertion times in the master equation (\ref{eq:WImaster-Delta}), we chose $y_0=T/2$ and $t=T/6$ rounded up to the closest integer\footnote{As discussed in Ref.~\cite{Bulava:2016ktf}, the temporal extent of our lattices is odd, so there is no central time-slice.}. The idea behind this choice is to place the operators  as far away from the temporal boundaries as possible, so as to suppress boundary induced cutoff effects, while keeping the individual operators apart from each other, thus avoiding contact terms.

In Fig. \ref{fig:chiral_extrapolation} we show the chiral extrapolation of our preferred determination WI($1468$), at $\beta=3.676$, where quark masses cover a large range in lattice units. We compare results obtained from the Ward identity with and without the mass term (i.e., the term with two pseudoscalar insertions in Eq.~(\ref{eq:SPImassintlatt})). We see that in the ``massive'' case our results display a linear behaviour in the whole mass range. In addition statistical uncertainties are smaller and the data show an almost flat dependence on $am$, resulting to a more reliable chiral extrapolation. Therefore, we obtain $Z_\mathrm{S}/Z_\mathrm{P}$ in the chiral limit by fitting linearly the results of the ``massive" case. For this fit we employ orthogonal distance regression \cite{Boggs1989} which takes into account not only errors in the dependent, but also in the independent variable. The error obtained from this procedure for the chirally extrapolated $Z_\mathrm{S}/Z_\mathrm{P}$ is in general larger compared to the one obtained from a standard least squares fit. Results for the individual ensembles as well as the chiral extrapolations are summarised in Table \ref{tab:zszp}, which will be discussed in Subsection~\ref{sec:scaling}.
\begin{table*}
\centering
\begin{tabular*}{\textwidth}{@{\extracolsep{\fill}}cl >{\bfseries}lllll@{}}
	\toprule
	ID   & \multicolumn{1}{c}{$am$}     & \multicolumn{1}{c}{WI($1468$)}   & \multicolumn{1}{c}{WI($1245$)}  & \multicolumn{1}{c}{$\mathrm{L}_1$}    & \multicolumn{1}{c}{WI($4488$)}   & \multicolumn{1}{c}{$\mathrm{L}_2$} \\
	\midrule
	A1k1 & $-$0.00282(62)           & 1.550(28)    & 1.554(46)  & 1.662(53)                    & 2.320(493)   & 1.688(76)                 \\
	A1k3 & \phantom{$-$}0.00127(91) & 1.513(50)    & 1.469(48)  & 1.863(86)                    & 1.439(814)   & 1.570(130)                \\
	A1k4 & $-$0.00113(34)           & 1.510(39)    & 1.519(62)  & 2.120(147)                   & 2.712(348)   & 1.685(47)                 \\
	& \phantom{$-$}0.0         & 1.514(32)    & 1.495(34)  & 1.863(83)                    & 2.244(453)   & 1.644(63)                 \\
	\midrule
	E1k1 & \phantom{$-$}0.00269(20) & 1.359(14)    & 1.337(16)  & 1.527(33)                    & 1.679(216)   & 1.450(39)                 \\
	E1k2 & $-$0.00017(17)           & 1.333(14)    & 1.323(17)  & 1.497(38)                    & 1.937(184)   & 1.452(32)                 \\
	& \phantom{$-$}0.0         & 1.334(13)    & 1.324(16)  & 1.498(36)                    & 1.922(175)   & 1.452(30)                 \\
	\midrule
	B1k1 & \phantom{$-$}0.00554(20) & 1.257(10)    & 1.259(14)  & 1.346(17)                    & 1.456(148)   & 1.267(26)                 \\
	B1k2 & \phantom{$-$}0.00444(31) & 1.249(17)    & 1.236(22)  & 1.352(29)                    & 1.088(242)   & 1.234(36)                 \\
	B1k3 & \phantom{$-$}0.00110(21) & 1.272(13)    & 1.272(14)  & 1.337(20)                    & 1.374(150)   & 1.314(32)                 \\
	 B1k4 & $-$0.00056(16)           & 1.250(9)     & 1.248(11)    & 1.312(24)                    & 1.667(162)   & 1.327(27)                 \\
	 & \phantom{$-$}0.0         & 1.255(8)     & 1.255(9)     & 1.323(17)                    & 1.528(117)   & 1.320(21)                 \\
	\midrule
	C1k1 & \phantom{$-$}0.01320(17) & 1.182(6)     & 1.176(7)   & 1.191(8)                     & 0.793(124)   & 1.150(32)                 \\
	C1k2 & \phantom{$-$}0.00601(12) & 1.174(7)     & 1.172(10)  & 1.200(12)                    & 1.250(140)   & 1.171(21)                 \\
	C1k3 & $-$0.00112(12)           & 1.178(11)    & 1.178(12)  & 1.198(17)                    & 1.190(129)   & 1.166(15)                 \\
	& \phantom{$-$}0.0         & 1.174(8)     & 1.176(10)  & 1.200(13)                    & 1.236(109)   & 1.167(13)                 \\
	\midrule
	D1k2 & \phantom{$-$}0.00074(22) & 1.145(25)    & 1.149(24)  & 1.157(20)                    & 1.202(322)   & 1.147(41)                 \\
	D1k4 & $-$0.00007(4)            & 1.143(2)     & 1.143(6)   & 1.148(6)                     & 1.147(31)    & 1.144(5)                  \\
	& \phantom{$-$}0.0         & 1.143(3)     & 1.144(5)   & 1.148(6)                     & 1.152(39)    & 1.144(6)                  \\
	\bottomrule
\end{tabular*}
\caption{%
Summary of results for $am$ and $Z_{\mathrm S}/Z_{\mathrm P}$ from different Ward identity determinations, labelled by WI($abcd$). The Ward identity linear combinations ${\rm L}_1$ and ${\rm L}_2$ are defined in Eqs. (\ref{eq:L1}) and (\ref{eq:L2}). In all Ward identities the mass terms with two pseudoscalar insertions in the bulk have been included; cf. eq. (\ref{eq:SPImassintlatt}). The errors quoted for the individual ensembles are statistical; the uncertainty on the values at the chiral point stem from the orthogonal distance regression procedure of Ref.~\cite{Boggs1989}. Our preferred $Z_{\mathrm S}/Z_{\mathrm P}$ estimates are obtained from the WI($1468$) results (in boldface).
}\label{tab:zszp}
\end{table*}

\subsection{Scaling}
\label{sec:scaling}
\begin{figure*}[t]
	\centering
	\includegraphics[width=0.85\linewidth]{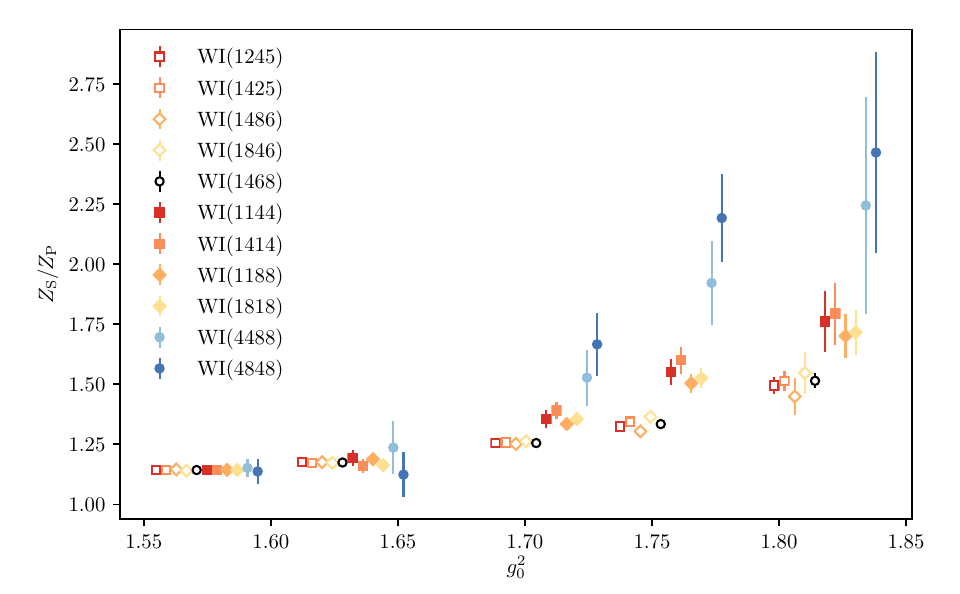}
	\caption{Dependence of $Z_\mathrm{S}/Z_\mathrm{P}$ on the gauge coupling $g_0^2$. Results are obtained from the $11$ Ward identity classes listed in Table~\ref{tab:F-classes}. Open symbols are used for the Ward identity classes with connected-quark diagrams only; closed symbols denote Ward identity classes with both connected- and disconnected-quark diagrams. Closely related Ward identities (which are separated by a single horizontal line in Table~\ref{tab:WIclasses}) are shown with the same symbol. Data from WI($1144$) are shown at their exact abscissa position, while the others have been slightly displaced in the $g_0^2$-direction, in order to improve visibility.}
	\label{fig:zszp_overview}
\end{figure*}
In Table~\ref{tab:F-classes} we have listed $11$ classes of distinct Ward identities; each of them is a different relation between correlation function differences $\Delta_k$ $(k=1,3,5,7,8)$ and $F_{\mathrm{S};1}$, from which $Z_\mathrm{S}/Z_\mathrm{P}$ may be obtained. In Fig.~\ref{fig:zszp_overview} we show these determinations in the chiral limit as functions of the gauge coupling $g_0^2$.  It is evident, as argued in Section \ref{sec:corr-funct}, that there are very strong correlations between results obtained on the same configuration ensembles from ``similar'' Ward identity classes, as grouped in Table~\ref{tab:WIclasses}.

We are thus led to select, from the plethora of Ward identities, four representative determinations of $Z_\mathrm{S}/Z_\mathrm{P}$. Two of these involve only quark connected diagrams. These are WI($1245$) and the linear combination $\mathrm{L}_1$, leading to Eq.~(\ref{eq:L1}). The other two determinations involve both quark connected and disconnected diagrams and are therefore numerically more challenging. Here we chose WI($4488$), and the linear combination $\mathrm{L}_2$, leading to Eq.~(\ref{eq:L2}). The results for each ensemble and in the chiral limit are shown in Table~\ref{tab:zszp}.

\begin{figure}[h]
	\centering
	\includegraphics[width=\linewidth]{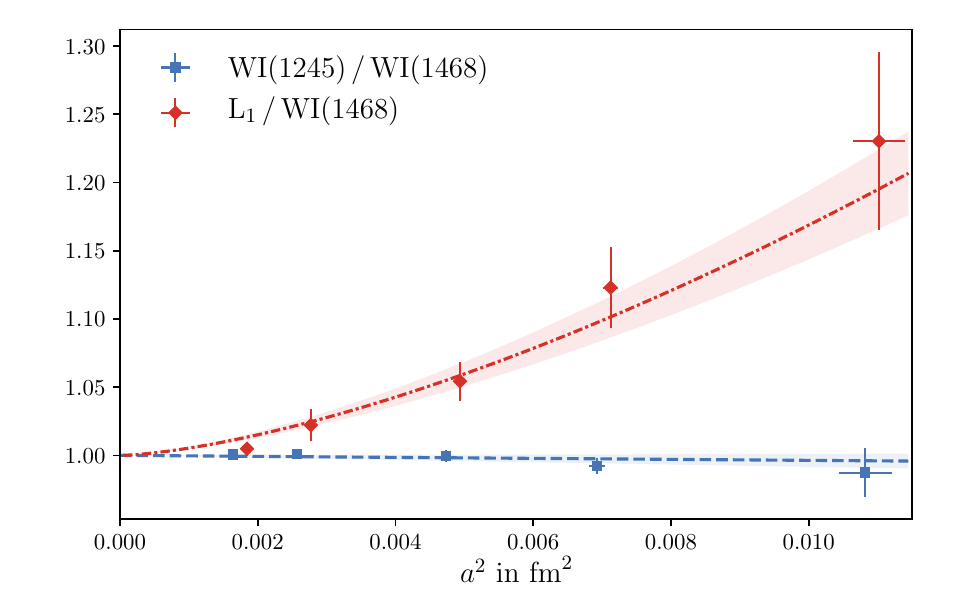}
	\includegraphics[width=\linewidth]{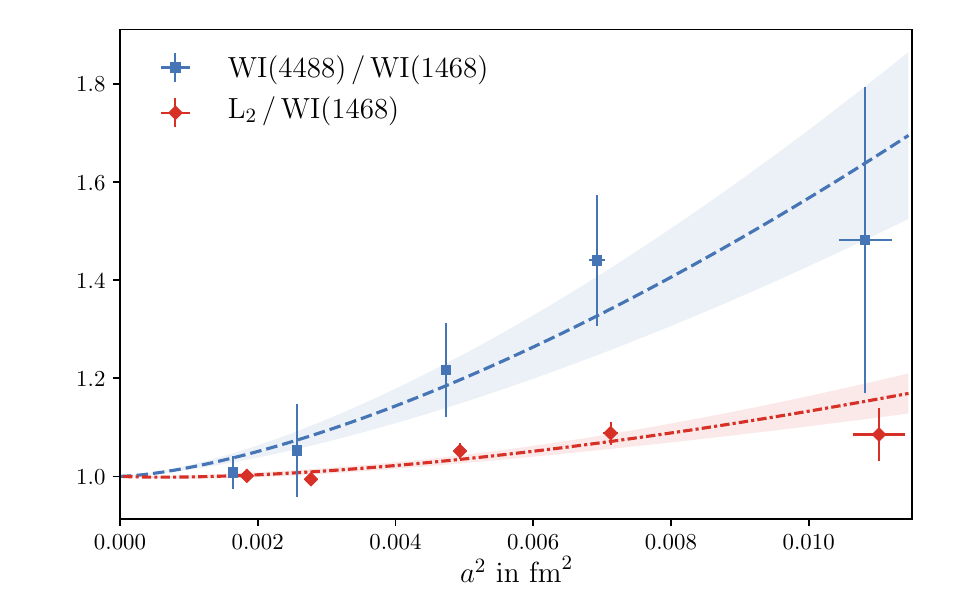}
	\caption{Lattice spacing dependence of the ratio of different $Z_{\mathrm S}/Z_{\mathrm P}$ determinations to $Z_{\mathrm S}/Z_{\mathrm P}$ from WI($1468$). The top panel depicts results from Ward identities which involve quark-connected diagrams only, while the bottom panel shows results from Ward identities which also involve quark-disconnected diagrams.}
	\label{fig:con_scaling}
\end{figure}

To evaluate the relative cutoff effects among our different results, we form ratios of $Z_\mathrm{S}/Z_\mathrm{P}$, obtained from each of the four determinations described above, to $Z_\mathrm{S}/Z_\mathrm{P}$ from our preferred identity WI($1468$). We investigate the lattice spacing dependence of each of these four ratios which, in our Symanzik-improved setup, consists of powers of $a^2$ and higher. The ratios are known to tend to unity in the continuum limit. We therefore fit them with polynomials in the lattice spacing, constrained to be $1$ at the origin. Results are displayed in Fig.~\ref{fig:con_scaling}. The top panel of the figure displays results from the first two determinations, without quark disconnected contributions.
The deviations from $1$ in the ratio WI($1245$)$/$WI($1468$) are very mild and can be described by a single term quadratic in the lattice spacing with $\chi^2/\text{d.o.f}=0.474$. For the ratio $\mathrm{L}_1/$WI($1468$) the deviation from $1$ as well as the statistical uncertainties are larger. A glance at Fig.~\ref{fig:con_scaling} should convince the reader that the data cannot be described by a single-parameter fit with a quadratic term. Fitting with $1 + c_2 a^2 + c_3 a^3$ results to $c_2 = -9.3(4.3)$, $c_3 = 303 (71)$ and $\chi^2/\text{d.o.f}=0.138$. A one-parameter fit with a term proportional to $a^3$ gives $c_3 = 169(22)$ with $\chi^2/\text{d.o.f}=0.775$; this is the curve shown in 
Fig.~\ref{fig:con_scaling}. The bottom panel of Fig.~\ref{fig:con_scaling} displays results from the determinations with quark disconnected contributions. Again it is obvious that none of the data displays a pure $a^2$-dependence. Fitting the ratio WI($4488$)$/$WI($1468$) with $1 + c_2 a^2 + c_3 a^3$ results to $c_2 = -26(28)$, $c_3 = 911 (410)$ and $\chi^2/\text{d.o.f}=0.494$; note that $c_2$ is compatible with zero. Fitting by $1 +  c_3 a^3$ gives $c_3 = 567(131)$ and $\chi^2/\text{d.o.f}=0.511$; this is the fit shown in the Figure. For the ratio $\mathrm{L}_2/$WI($1468$) we again fit with two parameters, one quadratic and one cubic in the lattice spacing, obtaining $c_2 = -7.8(4.6)$, $c_3 = 211(68)$ and $\chi^2/\text{d.o.f}=1.719$. The relatively large value for $\chi^2/\text{d.o.f}$ can be traced to the data point at the coarsest lattice spacing. All four cases conform with the theoretical expectation of $\mathrm{O}(a^2)$ ambiguities or higher. We did not find any evidence for $\mathrm{O}(a)$ cutoff effects; trying to fit an additional term proportional to $a$ gives coefficients which are zero within errors.

\subsection{Interpolation formula}
\label{sec:interpolation_formula}
\begin{figure*}[t]
	\centering
	\includegraphics[width=0.85\linewidth]{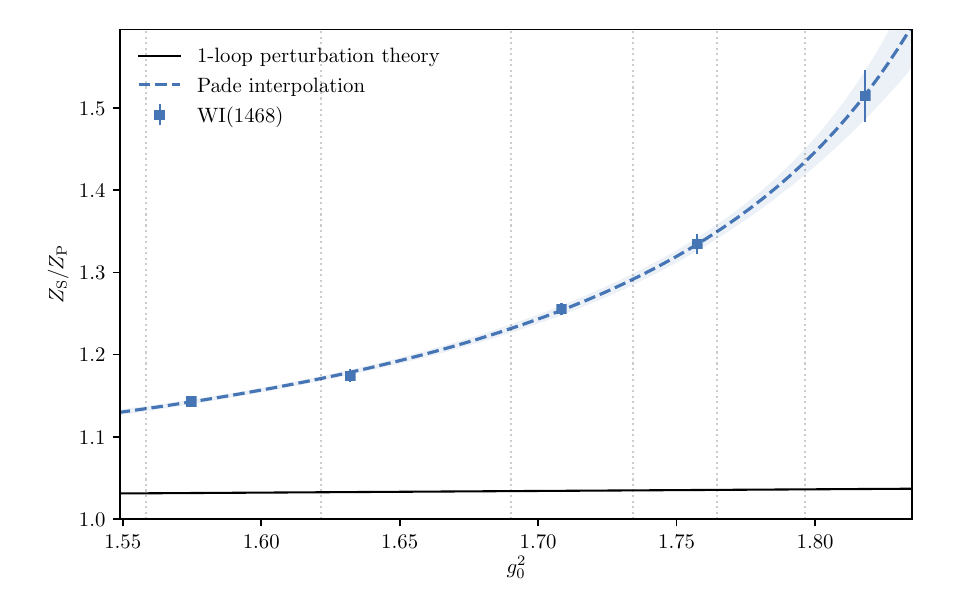}
	\caption{$Z_\mathrm{S}/Z_\mathrm{P}$ results from WI($1468$), extrapolated to the chiral point, plotted against the bare gauge coupling $g_0^2$. The Pad\'{e} interpolation formula (\ref{eq:pade_fit}), shown with errorband, is used to propagate the statistical uncertainty. The 1-loop perturbative result from Ref.~\cite{Constantinou:2009tr} is shown for comparison. The vertical dashed lines indicate the CLS couplings of Refs.~\cite{Bruno:2014jqa,Bali:2016umi,Mohler:2017wnb}.}
	\label{fig:zszp_fit}
\end{figure*}
To facilitate the use of our $Z_\mathrm{S}/Z_\mathrm{P}$ results in large volume simulations, we provide an interpolation formula for lattice spacings $0.04\,$fm$\,\lesssim a \lesssim 0.1\,$fm. 
Having tried several fit ansätze, we opt for a Pad\'{e} interpolation constrained by the 1-loop value \cite{Constantinou:2009tr} of the form
\begin{subequations}
\label{eq:pade_fit}
\begin{align}
\bigg(\frac{Z_\mathrm{S}}{Z_\mathrm{P}}\bigg)(g_0^2)=1+0.020164\,g_0^2\times \frac{1+Z_\mathrm{SP}^{(0)}g_0^2+Z_\mathrm{SP}^{(1)}g_0^4}{1+Z_\mathrm{SP}^{(2)}g_0^2}\,,
\\
Z_\mathrm{SP}^{(0)}=-0.5357\,,\quad Z_\mathrm{SP}^{(1)}=0.2883\,,\quad Z_\mathrm{SP}^{(2)}=-0.5117\,,
\end{align}
with the covariance matrix
	\begin{align}
	&\mathrm{cov}(Z_\mathrm{SP}^{(i)},Z_\mathrm{SP}^{(j)}) \nonumber \\
	=&{\scriptsize\begin{pmatrix}
	\begin{tabular}{@{}*{3}{S[table-format = +2.4e+1]}}
+2.0195e-01 & -1.3844e-01 & -4.1248e-03 \\
-1.3844e-01 & +9.5121e-02 & +2.8754e-03 \\
-4.1248e-03 & +2.8754e-03 & +9.6128e-05
	\end{tabular}
	\end{pmatrix}}\,,
	\end{align}
\end{subequations}
and $\chi^2/\text{d.o.f.}=0.169$.

As the functional form in the non-perturbative coupling region is in
principle unknown, we investigated the significance of systematic effects
by also experimenting with alternative forms of interpolating functions
(such as higher-order Pad\'{e}s, exponentials and polynomials), constrained to monotonically approach the 1-loop perturbation theory result.
However, among those describing our results reliably (as signaled by an
acceptable $\chi^2/\text{d.o.f.}$) practically coincide with the interpolation (\ref{eq:pade_fit}) in the fitted range of couplings, so that the associated systematic errors are negligible compared to the statistical ones. Therefore, we only account for systematic uncertainties when extrapolating with Eq. (\ref{eq:pade_fit}) to values slightly outside the fitted range by adding a systematic error of $50$\% of the size of the statistical one in quadrature. This prescription is applied at $\beta=3.85$, which corresponds to the finest lattice spacing simulated by the CLS effort. 

The WI($1468$) results with the interpolation are shown in Fig. \ref{fig:zszp_fit}, where they are also compared to the prediction of 1-loop perturbation theory. The vertical dashed lines mark the bare couplings used in CLS simulations, to which we want to interpolate our results.   Results for $Z_\mathrm{S}/Z_\mathrm{P}$ at the $g_0^2$-values used in $N_\mathrm{f} = 2 + 1$ CLS simulations are given in Table \ref{tab:cls}.
\begin{table}
\centering
\renewcommand{\arraystretch}{1.25}
\setlength{\tabcolsep}{3pt}
\begin{tabular}{llll}
	\toprule
	\multicolumn{1}{c}{$\beta$} & \multicolumn{1}{c}{WI($1468$)} & \multicolumn{1}{c}{\cite{deDivitiis:2019xla} LCP-0} & \multicolumn{1}{c}{\cite{deDivitiis:2019xla} LCP-1} \\
	\midrule
      3.85 & 1.1343(25)   & 1.1437(33)                        & 1.1441(24)                        \\
      3.7  & 1.1709(23)   & 1.2047(34)                        & 1.2023(25)                        \\
      3.55 & 1.2317(48)   & 1.3073(72)                        & 1.2971(51)                        \\
      3.46 & 1.2914(64)    & 1.409(10)                         & 1.3866(70)                        \\
      3.4  & 1.3497(83)   & 1.509(12)                         & 1.4720(77)                        \\
      3.34 & 1.435(15) &  1.662(19)                       &  1.595(11)                       \\
	\bottomrule
\end{tabular}
\caption{%
$Z_\mathrm{S}/Z_\mathrm{P}$ results from WI($1468$) (second column) and from Ref.~\cite{deDivitiis:2019xla} for two lines of constant physics (LCP), specified there. The inverse gauge couplings $\beta$ are those used in $N_\mathrm{f} = 2 + 1$ CLS simulations~\cite{Bruno:2014jqa,Bali:2016umi,Mohler:2017wnb}. The error of the WI($1468$) results is the statistical uncertainty propagated from the interpolation formula (\ref{eq:pade_fit}) except for $\beta=3.85$ where we added a systematic uncertainty, $50$\% of the size of the statistical one, in quadrature. For the results of the two LCP columns we combine the errors of $Z$ (from Ref.~\cite{deDivitiis:2019xla}) and $Z_\mathrm{A}$ (from Ref.~\cite{DallaBrida:2018tpn}) in quadrature.
}\label{tab:cls}
\end{table}

\subsection{Comparison with previous works}
\begin{figure*}[thb]
	\centering
	\includegraphics[width=0.85\linewidth]{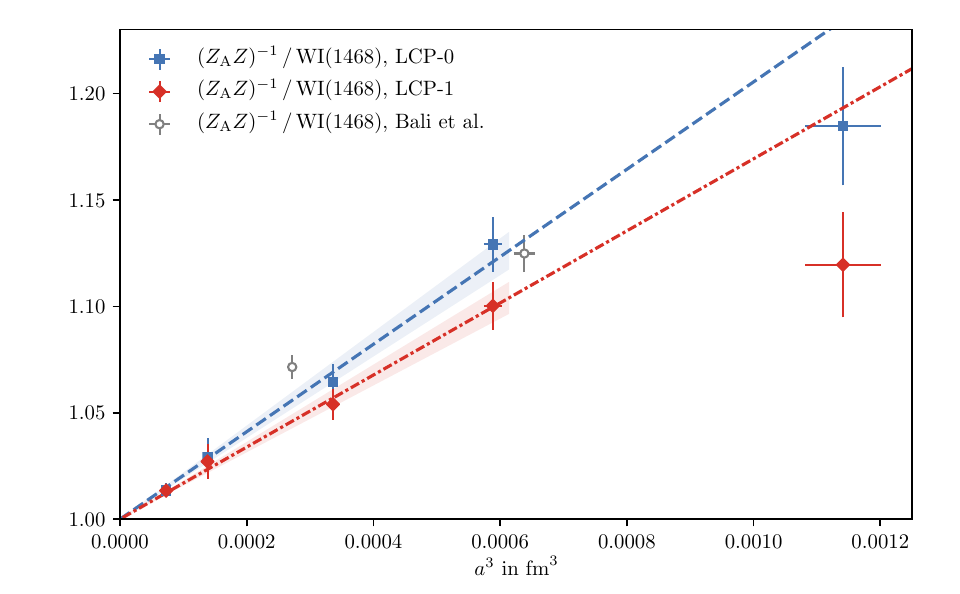}
	\caption{Scaling behaviour of the ratio of $Z_\mathrm{S}/Z_\mathrm{P}$ deduced  from results in Refs.~\cite{deDivitiis:2019xla,Bali:2016umi} to $Z_\mathrm{S}/Z_\mathrm{P}$ from WI($1468$).}
	\label{fig:Z_scaling}
\end{figure*}
We are not aware of any direct determinations of $Z_\mathrm{S}/Z_\mathrm{P}$ in our specific setup, but we can compare our findings, using existing results for the quark mass renormalisation constant $Z \equiv Z_\mathrm{P}/(Z_\mathrm{S} Z_\mathrm{A})$. The idea is to compute $Z_\mathrm{S}/Z_\mathrm{P}=(Z Z_\mathrm{A})^{-1}$, with $Z$ from either Ref.~\cite{Bali:2016umi} or Ref.~\cite{deDivitiis:2019xla}, and $Z_\mathrm{A}$ from Ref.~\cite{DallaBrida:2018tpn}. In Ref.~\cite{Bali:2016umi}, $Z$ has been computed on large-volume CLS ensembles, from the relation between PCAC quark masses $m_{ij}$ and subtracted quark masses $m_{\mathrm{q},ij}$ (see Section~\ref{section:qmasses} and \ref{app:renimp} for these mass definitions). The $Z$-results in Ref.~\cite{deDivitiis:2019xla} were obtained on almost the same gauge ensembles used in this work\footnote{We additionally use ensemble C1k1.} at small volumes and nearly-chiral sea quark masses. The method of Ref.~\cite{deDivitiis:2019xla} is based on suitable combinations of renormalised quark masses, defined both through the PCAC relation and the subtracted bare mass, evaluated in the $\mathrm{O}(a)$ improved theory with non-degenerate valence quarks, including all necessary counterterms. Results are quoted for two different lines of constant physics labeled LCP-0 and LCP-1, which differ by the values at which the quark masses in the valence sector are kept fixed as $g_0$ is varied. 

We compute the ratio of $1/(Z Z_\mathrm{A})$ from Refs.~\cite{Bali:2016umi} and \cite{deDivitiis:2019xla} to $Z_\mathrm{S}/Z_\mathrm{P}$ from our preferred WI($1468$). We investigate the lattice spacing dependence of this ratio, which consists of powers of $a^2$ and higher, and tends to unity in the continuum limit. The results are plotted in Fig.~\ref{fig:Z_scaling}. Polynomial fits are performed on the LCP-0 and LCP-1 ratios, excluding the data of the coarsest ensembles, which display poor scaling behaviour and large errors. A two-parameter fit of the form $1+ c_2 a^2 + c_3 a^3$ results to $\chi^2/\text{d.o.f} = 0.281$, $c_2=-2.5(3.7)$ and $c_3=242(57)$ for LCP-0, and $\chi^2/\text{d.o.f} = 0.166$, $c_2=1.5(2.8)$ and $c_3=148(45)$ for LCP-1, in both cases $c_2$ is consistent with zero. We thus prefer to plot the results as functions of $a^3$ in Fig. \ref{fig:Z_scaling}, where we also show a one-parameter fit of the form $1+ c_3 a^3$; for this ansatz we obtain $\chi^2/\text{d.o.f} = 0.300$, $c_3 = 206(14)$ for LCP-0 and $\chi^2/\text{d.o.f} = 0.170$, $c_3 = 169(12)$ for LCP-1\footnote{Since we neglect correlations between our results and those of Ref.~\cite{deDivitiis:2019xla}, the error in their ratio is probably overestimated. This explains the small values of $\chi^2/\text{d.o.f}$.}. We interpret this as confirmation that the two methods are compatible w.r.t. the expected lattice spacing ambiguities and that the effects of $\rmO(a^2)$ are sub-dominant compared to the next higher order.

Let us briefly comment on the possible benefits of the respective results
on $Z_\mathrm{S}/Z_\mathrm{P}$ collected in Table~\ref{tab:cls}, originating from the different approaches underlying Ref.~\cite{deDivitiis:2019xla} and this work.
First, one observes comparable uncertainties between the two.
While the method of that reference involves combinations of simpler and thus typically less noisy correlation functions (i.e., with only one operator insertion in the bulk) as well as an accurate computation of the valence quark mass dependence prior to the chiral extrapolations, our estimates on $Z_\mathrm{S}/Z_\mathrm{P}$ from the more direct Ward identity
approach followed here exhibit an overall flatter and, at larger
couplings, less steep $g_0^2$-dependence. This points to generically smaller cutoff effects so that continuum extrapolations of quantities where it enters may be expected to become better controlled and more precise in the long run, because they are also less affected by unpleasantly significant admixtures of higher-order cutoff effects. 

The results for $Z$ presented in \cite{Bali:2016umi}, stemming from large-volume calculations on a subset of the CLS ensembles, are only available at two values of the bare coupling, which do not coincide with the
couplings investigated in this work. In order to compare with our results we make use of the interpolation formula Eq.~(\ref{eq:pade_fit}).  Although the estimates for $Z$ from Ref.~\cite{Bali:2016umi} are only available at two values of the bare coupling and we hence do not attempt a fit in this case, we notice that they are compatible with LCP-0. 

In summary, comparison with earlier works is consistent with the
expectation that all ambiguities between different determinations of
$Z_\mathrm{S}/Z_\mathrm{P}$ show a scaling according to $\rmO(a^2)$ or higher.
However, the size of these ambiguities is quite large and may still have
a relevant impact on applications as described in the next Section.
\section[Application: quark mass computations with Wilson fermions]{Application: quark mass computations with\\Wilson fermions}
\label{section:qmasses}
We will now discuss a method of computing quark masses with Wilson fermions which uses  the ratio $Z_\mathrm{S}/Z_\mathrm{P}$.

First we review the well-established ``PCAC quark mass method''. It is the conventional ALPHA Collabora\-tion approach, which relies on the PCAC definition of quark 
masses $m_{ij}$ of Eq.~(\ref{eq:PCACmassij}). These bare current masses are computed on large physical volumes\footnote{The ALPHA Collaboration 
has performed these calculations for quenched QCD with Schrödinger functional boundary
conditions; see Ref.~\cite{Garden:1999fg}. The CLS effort determined quark masses for $\NF=2$ QCD with periodic boundary conditions \cite{Fritzsch:2012wq,Heitger:2013oaa} and for
$\NF=2+1$ QCD with open boundary conditions~\cite{Bruno:2019vup,Heitger:2019ioq}.} and for a range of couplings typical of hadronic,
low-energy scales $\mu_{\mathrm{had}} \sim \Lambda_{\mathrm{QCD}}$. Although we keep our notation as general as possible, for concreteness we consider
a theory with $\NF = 2+1$ dynamical fermions; i.e.\ the two lightest flavours are degenerate in
mass while the third flavour is heavier ($m_{\rm q,1} = m_{\rm q,2} < m_{\rm q,3}$).

We see from Eq.~(\ref{eq:renmassPCAC}) that the renormalised light mass is given by
\begin{align}
\label{eq:renmassPCAC12}
m_{1,\rm R} & = m_{2,\rm R} 
=  \dfrac{Z_\mathrm{A}}{Z_\mathrm{P}} \,\, m_{12} \, \times \\
\Big [ 1 \, & + (\ba-\bp) a m_{\mathrm{q},12} \, + \, (\bar b_\mathrm{A}- \bar b_\mathrm{P})a \Tr(M_\mathrm{q}) \Big ]  + \rmO(a^2) \,. \nonumber
\end{align}
The ratio of the heavy to light renormalised masses is also derived from the above expression:
\begin{align}
\label{eq:massrat}
&\dfrac{m_{3,\rm R}}{m_{1,\rm R}}  \\
=&  2 \, \dfrac{m_{13}}{m_{12}} \, \Big [ 1 + (\ba - \bp) \dfrac{(am_{\rm q,3} - am_{\rm q,2} )}{2} \, \Big ]  -  1 + \rmO(a^2) \,. \nonumber
\end{align}
Knowing the renormalised light mass from Eq.~(\ref{eq:renmassPCAC12}), and the ratio of the heavy and light renormalised masses from Eq.~(\ref{eq:massrat}), the up/down and strange masses are obtained~\cite{Bruno:2016plf,Bruno:2019vup}. So {\it in principle} this method requires:
\begin{enumerate}
\item
The axial current normalisation $Z_\mathrm{A}(g_0^2)$ and the renormalisation constant $Z_\mathrm{P}(g_0^2,\mu_{\mathrm{had}})$ of the non-singlet pseudoscalar density; the latter carries the renormalisation scheme and scale dependence of the continuum quark mass. In our $\NF=3$ setup, these may be found in Refs.~\cite{DallaBrida:2018tpn} and \cite{Campos:2018ahf}, respectively.
\item
The Symanzik-improvement coefficients $(\ba - \bp)$ and $(\bar b_\mathrm{A} - \bar b_\mathrm{P})$. Non-perturbative $(\ba - \bp)$-estimates in our setup may be found in Ref.~\cite{deDivitiis:2019xla}. Note that in perturbation theory $(\bar b_\mathrm{A} - \bar b_\mathrm{P}) \sim {\rmO}(g_0^4)$, so that the term proportional to this coefficient is habitually dropped.
\item
It is also noteworthy that Eq.~(\ref{eq:massrat}) does not require knowledge of $\kappa_\mathrm{crit}$, which is however needed in $m_{\mathrm{q},12}$ and $\Tr(M_\mathrm{q})$ in Eq.~(\ref{eq:renmassPCAC12}). We shall return to this point in Subsection \ref{eq:mass_improvement}.
\end{enumerate}

Based on the results of Ref.~\cite{Bhattacharya:2005rb} for Symanzik-im\-proved quark masses with Wilson fermions, an alternative approach, known as the ``ratio-difference method'', has been proposed in Ref.~\cite{Durr:2010aw}. The renormalised quark mass difference is given by
\begin{gather}
\begin{split}
\label{eq:massdiff}
m_{3,\rm R} &- m_{1,\rm R} =  Z_{\mathrm S}^{-1}  \Big [ m_{\mathrm{q},3} - m_{\mathrm{q},1} \Big ]  \,\, \times \\
& \Big [ 1 + a 2 b_m \, m_{\mathrm{q},13}
+ a \bar b_m \Tr(M_\mathrm{q}) \Big ] +  \rmO(a^2) \,.
\end{split}
\end{gather}
Knowing the renormalised mass difference from Eq.~(\ref{eq:massdiff}), and the ratio of the heavy and light renormalised masses from Eq.~(\ref{eq:massrat}), the up/down and strange masses are obtained. So {\it in principle} this method requires:
\begin{enumerate}
\item
The renormalisation constant $Z_\mathrm{S}(g_0^2,\mu_{\mathrm{had}})$ of the non-singlet scalar density, which carries the renormalisation scheme and scale dependence of the continuum quark mass.
\item
The Symanzik-improvement coefficients $(\ba - \bp)$, $b_m$ and $\bar b_m$. Non-perturbative estimates of the $b_m$-coefficient in this setup may be found in Ref.~\cite{deDivitiis:2019xla}.\footnote{In perturbation theory $2b_m = -1 + {\rmO}(g_0^2)$ and the non-perturbative estimates of Ref.~\cite{deDivitiis:2019xla} are also numerically sizeable. Thus this Symanzik counterterm is expected to remove large $\rmO(a)$ effects, especially in future computations of heavy flavour quark masses (charm etc.).} Since $\bar b_m \sim {\rmO}(g_0^4)$, the term proportional to  $\Tr(M_\mathrm{q})$ is habitually dropped.
\item
The critical hopping parameter $\kappa_\mathrm{crit}$ is needed in $m_{\mathrm{q},13}$ and $\Tr(M_\mathrm{q})$ in Eq.~(\ref{eq:massdiff}). We shall return to this point in Subsection \ref{eq:mass_improvement}.
\end{enumerate}

We have outlined the basic idea behind the PCAC quark mass method and the ratio-difference method, listing the renormalisation parameters and improvement coefficients required by each one. The most crucial difference is that in the PCAC quark mass method all bare masses are given in terms of the current masses $m_{12}$ and $m_{13}$, which
are renormalised by $Z_\mathrm{P}^{-1} Z_\mathrm{A}$, while in the ratio-difference method the bare mass difference is the exactly known $[m_{\mathrm{q},3} - m_{\mathrm{q},1}]$, which is renormalised by $Z_\mathrm{S}^{-1}$. It is not possible to determine $Z_{\mathrm S}$ with a Schr\"odinger functional renormalisation condition analogous to that introduced in Ref.~\cite{Capitani:1998mq} for $Z_{\mathrm P}$. The latter involves correlation functions with a pseudoscalar source at the boundary (see Eq.~(\ref{eq:boundary-sources})) and the pseudoscalar scalar operator at the bulk. If we place a scalar operator at the bulk, keeping the pseudoscalar boundary source, the correlation function vanishes due to parity. Nor is it possible to have a scalar source at the boundary and the scalar density at the bulk, since this would result in the product $P_+ P_-$ of the projection operators of the boundary quarks and the vanishing of the correlation function. An option would be to impose a renormalisation condition on the correlation function $\langle  \cO^{\prime a} \, S^b(x) \, \cO^c \rangle$, with the two pseudoscalar boundary sources $\cO^{\prime a}$ and $\cO^c$ and the scalar operator $S^b$ in the bulk. This would be an acceptable intermediate scheme of the Schr\"odinger functional variety, but different than the one introduced in Ref.~\cite{Capitani:1998mq} for $Z_{\mathrm P}$. Thus, the renormalised quark masses $m_{1\rm R},m_{3\rm R}$ obtained by combining Eqs.~(\ref{eq:renmassPCAC12}) and (\ref{eq:massrat}) (PCAC quark mass method with $Z_{\mathrm P}$) would be in a different scheme than those obtained from Eqs.~(\ref{eq:massdiff}) and (\ref{eq:massrat}) (difference-ratio method with $Z_{\mathrm S}$). Only results obtained for the scheme-independent renormalisation group invariant (RGI) masses from the two methods would be comparable. This comparison would be very useful but cumbersome, as it requires the computation from scratch of the step scaling function in the new intermediate scheme, from ratios of $Z_{\mathrm S}$'s at fixed renormalised coupling and two different renormalisation scales, and for a range of couplings. 

Given the above considerations, we are led to define the scalar operator renormalisation parameter through:
\begin{align}
\label{eq:ZS-SFscheme}
Z_{\mathrm S}(g_0^2,\mu_{\mathrm{had}}) =  \Bigg [ \dfrac{Z_{\mathrm S}(g_0^2,\mu_{\mathrm{had}})}{Z_{\mathrm P}(g_0^2,\mu_{\mathrm{had}})} \Bigg ] Z_{\mathrm P}(g_0^2,\mu_{\mathrm{had}}) \,.
\end{align}
This is our definition of the Schr\"odinger functional renormalisation scheme for the scalar non-singlet operator. The $Z_{\mathrm S}/Z_{\mathrm P}$-ratio on the r.h.s.\ is scale independent, being determined from Ward identities. Clearly, scalar and pseudoscalar densities have the same renormalisation group running properties (i.e., the same anomalous dimensions, the same step scaling functions in the continuum, etc.). So knowledge of the $Z_{\mathrm S}/Z_{\mathrm P}$ ratio enables us to obtain the light and  heavy quark masses in the usual Schr\"odinger functional scheme~\cite{Capitani:1998mq}, but with a different method based on mass differences (and $Z_{\mathrm S}$) combined with scale-independent PCAC mass ratios. The novel renormalisation and improvement patterns provide an important handle for the control and reduction of systematic effects related to the non-perturbative determination of renormalisation parameters and discretisation errors\footnote{This could be crucial in computations of heavier quark masses (charm etc.), where the discretisation errors become dominant.}. What is common in both methods is the renormalisation group running that takes us non-perturbatively from renormalised masses at low energy scales $\mu_{\mathrm{had}}$ to masses at large, perturbative scales $\mu_\mathrm{PT} \sim M_{\mathrm W}$, as described in Ref.~\cite{Capitani:1998mq}. For recent results on the running of quark masses in $\NF = 3$ QCD see Ref.~\cite{Campos:2018ahf}.

\subsection{Subtracted masses, PCAC masses, and redefined Symanzik counterterms}
\label{eq:mass_improvement}
We will close this section by reviewing how, in both methods, we can circumvent the need to use $\kappa_\mathrm{crit}$ in the Symanzik counterterms of Eqs.~(\ref{eq:renmassPCAC12}) and (\ref{eq:massdiff}), which feature subtracted masses $am_{\mathrm{q},ij}$ and $\Tr [a\Mq]$. This can be avoided by substituting these subtracted masses with current quark masses. Their relation is given by~\cite{Bhattacharya:2005rb},
\begin{equation}
\label{eq:mij_mqij}
    m_{ij} = Z  \bigg[ m_{\mathrm{q},ij} + \left( r_m - 1 \right) \dfrac{\Tr[\Mq]}{\NF} \bigg]  + \rmO(a) \,,
\end{equation}
where $Z(g_0^2)\equiv Z_{\mathrm P}/(Z_{\mathrm S} Z_{\mathrm A})$ and $r_m \equiv Z_{\mathrm S}/Z_{\mathrm S^0}$ are finite normalisations ($Z_{\mathrm S^0}$ is the renormalisation parameter of the singlet scalar density).
In the above we neglect $\rmO(a)$ terms, as they only contribute to $\rmO(a^2)$ in the $b$-counterterms of Eqs.~(\ref{eq:renmassPCAC12}) and (\ref{eq:massdiff}).
Substituting $am_{\mathrm{q},ij}\to am_{ij}$ in these expressions, we obtain respectively
\begin{align}
\begin{split}
m_{1,\rm R} & = m_{2\rm R} = \dfrac{Z_{\mathrm A}}{Z_{\mathrm P}} \, m_{12} \,\, \times \\
& \Bigg [ 1 + (\tilde b_\mathrm{A}-\tilde b_\mathrm{P}) am_{12}
+ \Bigg \{ (\tilde b_\mathrm{A}- \tilde b_\mathrm{P}) \dfrac{1-r_m}{r_m} \\
& + (\bar b_\mathrm{A} - \bar b_\mathrm{P}) \dfrac{\NF}{Z r_m} \Bigg \} \dfrac{a M_{\rm sum}}{\NF} \Bigg ] +  \rmO(a^2) \,,
\label{eq:renmassPCAC2}
\end{split}
\end{align}
and
\begin{align}
m_{3,\rm R} & - m_{1,\rm R} = Z_{\mathrm S}^{-1} \Big [ m_{\mathrm{q},3} - m_{\mathrm{q},1} \Big ]  \,\, \times \label{eq:massdiff2}\\
& \Bigg [ 1 + 2 \tilde b_m \, a m_{13} \nonumber\\
& + \Bigg \{  2 \tilde b_m \dfrac{1-r_m}{r_m} + \bar b_m \dfrac{\NF}{Z r_m} \Bigg \} \dfrac{a M_{\rm sum}}{\NF} \Bigg ] +  \rmO(a^2) \,,\nonumber
\end{align}
where we define
\begin{align}
\tilde b_\mathrm{A}- \tilde b_\mathrm{P} & \equiv  \dfrac{\ba - \bp}{Z} \,, \qquad  \tilde b_m  \equiv  \dfrac{b_m}{Z} \,, \\
\begin{split}
M_{\rm sum} & \equiv  m_{12} + m_{23} + \cdots + m_{(\NF-1)\NF} + m_{\NF 1} \\
&=  Z r_m \Tr[\Mq] + \rmO(a) \,. 
\label{eq:Msum}
\end{split}
\end{align}
Thus, $am_{\mathrm{q},ij}$ and $\kappa_\mathrm{crit}$ in Eqs.~(\ref{eq:renmassPCAC12}) and (\ref{eq:massdiff}) have been traded off for $m_{ij}$, $Z$, and $r_m$. Accurate non-perturbative estimates of $Z$, $(\ba-\bp)$, and $b_m$ in our $\NF=3$ setup have been reported in Ref.~\cite{deDivitiis:2019xla}. The term multiplying $M_{\rm sum}$ contains $(1-\r_m)/r_m$ and $(\bar b_\mathrm{A} - \bar b_\mathrm{P})$. To leading order in perturbation theory $r_m = 1 + 0.001158\,\CF\,\NF\,g_0^4$~\cite{Constantinou:2014rka,Bali:2016umi}; thus $(1-r_m)/r_m \sim  \rmO(g_0^4)$.  A first non-perturbative study of the coefficients $\bar b_\mathrm{A}$, $\bar b_\mathrm{P}$, and $\bar b_m$ produced noisy results with 100\% errors~\cite{Korcyl:2016ugy}. Since in perturbation theory $(\bar b_\mathrm{A}- \bar b_\mathrm{P}), \bar b_m \sim \rmO(g_0^4)$~\cite{Bhattacharya:2005rb}, the terms proportional to $M_{\rm sum}$ are habitually dropped. 

For completeness we also discuss a slightly different way to write the $b_m$-counterterm of the renormalised quark mass difference of Eq.~(\ref{eq:massdiff}), in close analogy to what is done in Ref.~\cite{Durr:2010aw}. The term in question is written as follows:

{\small
\begin{align}
\begin{split}
\label{eq:b_m-BMW}
&a b_m [ m_{\mathrm{q},3} + m_{\mathrm{q},1} ] =  a b_m [ m_{\mathrm{q},3} + m_{\mathrm{q},1} ] \Bigg [ \dfrac{m_{\mathrm{q},3} - m_{\mathrm{q},1}}{ m_{\mathrm{q},3} - m_{\mathrm{q},1}} \Bigg ]  \\
&= a b_m  \bigg [ m_{\mathrm{q},3} - m_{\mathrm{q},1} \bigg ] \dfrac{\bigg [\dfrac{m_{33^\prime}}{m_{12}} + 1 \bigg ] + \dfrac{2(1-r_m)}{r_m} \dfrac{M_{\rm sum}}{m_{12}N_\mathrm{f}}}{\bigg [\dfrac{m_{33^\prime}}{m_{12}} - 1 \bigg ]} \,.
\end{split}
\end{align}
}
We arrive at the second expression using Eq.~(\ref{eq:mij_mqij}) and introducing the PCAC mass $m_{33^\prime}$, which consists of two degenerate but distinct heavy valence flavours. Neglecting the term proportional to $M_{\rm sum}$ in Eq.~(\ref{eq:b_m-BMW}), we conclude that in this approximation the difference-ratio method is based on Eqs.~(\ref{eq:massrat}) and (\ref{eq:massdiff}), which depend on the exactly known subtracted quark mass difference $[m_{\mathrm{q},3} - m_{\mathrm{q},1}]$ and suitable PCAC quark mass ratios, but not on subtracted quark mass averages  $m_{\mathrm{q},ij}$ and $\kappa_\mathrm{crit}$.

\section{Conclusions}
In the present study we have addressed, for the first time within the
finite-volume Schrödinger functional setup, the non-perturbative
determination of the ratio of the scalar to pseudoscalar non-singlet
renormalisation constants $Z_\mathrm{S}/Z_\mathrm{P}$ in Wilson's lattice QCD, exploiting
suitable massive chiral Ward identities.
We have shown that in lattice QCD with three flavours of Wilson-Clover quarks
(with non-perturbative $\csw$~\cite{Bulava:2013cta}) and tree-level
Symanzik-improved gauge action, the Ward identities
are restored up to $\rmO(a^2)$ at finite lattice spacing.
In order to ensure a smooth dependence of the renormalisation constant
ratio on the bare gauge coupling, we have enforced a constant physics
condition by working with an approximately fixed physical volume of
spatial extent $L \approx 1.2\,\fm$ and $T/L \approx 3/2$.

Our main results are the parameterisation of $Z_\mathrm{S}/Z_\mathrm{P}$ in Eq.~(\ref{eq:pade_fit}),
valid for bare couplings $1.55\lesssim g_0^2\lesssim 1.85$
(i.e., lattice spacings $0.042\,\fm \lesssim a\lesssim 0.105\,\fm$),
as well as the values for $Z_\mathrm{S}/Z_\mathrm{P}$, given in Table~\ref{tab:cls}, at the bare couplings
typically employed in the large-volume $N_\mathrm{f}=2+1$ CLS
ensembles~\cite{Bruno:2014jqa,Bruno:2016plf,Bali:2016umi,Mohler:2017wnb}.
On the technical level, we had to treat properly the topology freezing encountered in our simulations, principally at the finest lattice
spacing, which may prevent a trustworthy estimation of the statistical error.
The operator character of Ward
identities ensures their validity in sectors of fixed topological
charge. Thus we have projected the correlation functions entering the
Ward identities onto the trivial topological sector
throughout our analysis.

Several checks have been performed, in order to guarantee the stability of the analysis and a careful
assessment of the statistical as well as the systematic errors. In particular, we have verified that results on $Z_\mathrm{S}/Z_\mathrm{P}$ from
the different classes of Ward identities at our disposal are perfectly
consistent with each other as expected, i.e., up to ambiguities of $\mathrm{O}(a^2)$ or even higher.
Among the various estimators for $[Z_\mathrm{S}/Z_\mathrm{P}](g_0^2)$,
our preferred choice, advocated in Eq.~(\ref{eq:pade_fit}), was guided by
the structural simplicity of the underlying chiral Ward identity, its
numerical precision, and its robustness against systematic effects.

Since the range of couplings covered in
this work matches those of the large-volume gauge field configurations
generated by CLS with the same lattice action, our result for $[Z_\mathrm{S}/Z_\mathrm{P}](g_0^2)$, combined with the scale dependent renormalisation factor $Z_\mathrm{P}$ from
\cite{Campos:2018ahf}, can be used in the computation of quark masses
as outlined in Section~\ref{section:qmasses}.
Work in this direction, extending the $(2+1)$-flavour computations of
light, strange and charm quark masses on the CLS ensembles reported in
refs.~\cite{Bruno:2019vup,Heitger:2019ioq}, is in progress.
\begin{acknowledgement}%
	We thank Stefan Sint, Christian Wittemeier, Carl Christian K\"oster and Simon Kuberski for helpful discussions and especially Carl for his valuable contributions in extending the set of ensembles used in our computations. A. V. wishes
to thank the Particle Physics Theory Group at WWU Münster and Trinity College Dublin for their hospitality. This work is supported by the Deutsche Forschungsgemeinschaft (DFG) through the Research Training Group \textit{GRK 2149: Strong and Weak Interactions -- from Hadrons to Dark Matter} (F. J. and J. H.). We acknowledge the computer resources provided by the Zentrum f\"ur Informationsverarbeitung of the University of M\"unster (PALMA \& PALMA II HPC clusters) and thank its staff for support.

\end{acknowledgement}

\numberwithin{equation}{section}
\begin{appendix}
	\section{Basic definitions}
\label{app:general}

We define non-singlet vector and axial vector currents in QCD with $\NF$ quarks as
\begin{align}
V_\mu^a(x)  &=  \mathrm{i} \bar \psi(x) \gamma_\mu T^a \psi(x) \,, \\
A_\mu^a(x)  &=  \mathrm{i} \bar \psi(x) \gamma_\mu \gamma_5 T^a \psi(x) \,,
\end{align}
with $a=1,\dots,(\NF^2-1)$ an $SU(\NF)$ flavour index. See \ref{app:sun} for our conventions regarding $SU(\NF)$ groups and $su(\NF)$ Lie algebras.
Analogously, non-singlet scalar and pseudoscalar densities are given by
\begin{align}
S^a(x) &= \mathrm{i} \bar \psi(x)  T^a \psi(x) \,,\\
P^a(x) &= \mathrm{i} \bar \psi(x) \gamma_5 T^a \psi(x) \,.
\end{align}

Axial transformations of the fermion fields are defined as:
\begin{gather}
\begin{split}
\psi(x) & \to  \psi^\prime(x) =  \exp\Big [ \mathrm{i} \epsilon^a(x) T^a \gamma_5 \Big ] \psi(x) \,, \\
\bar \psi(x) & \to \bar \psi^\prime(x)   =  \bar \psi(x) \exp \Big [ \mathrm{i} \epsilon^a(x) T^a \gamma_5 \Big ] \,.
\label{eq:axialtransf}
\end{split}
\end{gather}
Small axial field variations are obtained by expanding the above up to $\rmO(\epsilon)$:
\begin{gather}
\begin{split}
\delta_\mathrm{A} \psi(x) & =  \epsilon^a(x) \delta_\mathrm{A}^a \psi(x)  \approx  \mathrm{i} \epsilon^a(x) T^a \gamma_5 \psi(x) \, , \\
\delta_\mathrm{A} \bar \psi(x) & = \epsilon^a(x) \delta_\mathrm{A}^a \bar \psi(x) \approx   \mathrm{i} \epsilon^a(x) \bar \psi(x) T^a \gamma_5 \,.
\label{eq:axialtransinft}
\end{split}
\end{gather}
Note that in general these transformations are defined to be {\it local} (i.e., $\epsilon^a$ depends on space-time). Their {\it global} counterparts are related to  symmetries of the continuum theory (vector and chiral). 

In the Schrödinger functional framework, standard zero-momentum sources are defined as follows\footnote{In practice, instead of the sources ${\cal O}^a$ and ${\cal O}^{\prime a}$ defined in Eq.~(\ref{eq:boundary-sources}), we use pseudoscalar smeared sources with wavefunctions at the boundaries, as explained in \cite{Bulava:2015bxa}.}:
\begin{gather}
\begin{split}
\cO^a  &\equiv  \mathrm{i} a^6 \sum_{{\bf u},{\bf v}} \,\, \bar \zeta({\bf u}) \gamma_5 T^a \zeta({\bf v}) \,, \\
\cO^{\prime a}  &\equiv  \mathrm{i} a^6 \sum_{{\bf u^\prime},{\bf v^\prime}} \bar \zeta^\prime({\bf u^\prime}) \gamma_5 T^a \zeta^\prime({\bf v^\prime}) \,,
\label{eq:boundary-sources}
\end{split}
\end{gather}
where $\zeta$ and $\zeta^\prime$ are the quark fields at the Schr\"odinger functional boundaries $x_0$ = 0 and $x_0 = T$ , respectively.

	\section{Properties of \texorpdfstring{$su(\nf)$}{su(Nf)} Lie algebra generators}
\label{app:sun}
Our conventions for the $su(\nf)$ Lie Algebra are those of Appendix A.3. of Ref.~\cite{Luscher:1996sc}. In general, the anti-Hermitean generators of the algebra satisfy
\begin{equation}
\big [ T^a ,T^b \big ]   =  f^{abc} T^c \,.
\end{equation}
We work in the fundamental representation, with the generators normalised so that
\begin{eqnarray}
\label{eq:TTnorm}
\Tr \big[ T^a T^b\big ]  =  -\dfrac{1}{2} \delta^{ab} \,.
\end{eqnarray}
The anticommutator of these generators is given by
\begin{eqnarray}
\big \{ T^a ,T^b \big \}   =  -\mathrm{i} d^{abc} T^c - \dfrac{\delta^{ab}}{\nf} I_{\nf} \,,
\end{eqnarray}
where $I_{\nf}$ is the dimension-$\NF$ unit matrix.
The structure constants $f^{abc}$ are real and totally antisymmetric tensors, while $d^{abc}$ are real and totally symmetric.
Two useful identities are
\begin{equation}
\Tr [T^a T^b T^c ] = \dfrac{1}{4} \big [ \mathrm{i} d^{abc} - f^{abc} \big ] \,,
\label{eq:trTTT}
\end{equation}
\begin{align}
\Tr [T^a T^b T^c T^d] =& \dfrac{1}{4 \nf} \delta^{ab} \delta^{cd} \nonumber\\
&+  \dfrac{1}{8}  \big [d^{abe} + \mathrm{i} f^{abe} \big ] \, \big [d^{cde} + \mathrm{i} f^{cde} \big ]\nonumber \\
=& \dfrac{1}{8} \Big \{ \dfrac{2}{\Nf} \delta^{ab} \delta^{cd} + d^{abe} d^{cde} - f^{abe} f^{cde} \nonumber\\
&+ \mathrm{i} \big [d^{abe}  f^{cde} +  d^{cde} f^{abe} \big ] \, \Big \}\,.
\label{eq:trTTTT}
\end{align}

For $\nf = 2$ we have $T^a = \tau^a/(2\mathrm{i})$ ($\tau^a$ are the Pauli matrices), $f^{abc} = \epsilon^{abc}$ (the Levi-Civita symbol) and $d^{abc} = 0$. 

For $\nf = 3$ we have $T^a = \lambda^a/(2\mathrm{i})$ ($\lambda^a$ are the Gell-Mann matrices). The non-vanishing structure constants are
\begin{gather}
\begin{split}
f^{123} &= 1 \,, \\
f^{147} &= f^{246} = f^{257} = f^{345} = \dfrac{1}{2} \,, \\
f^{156} &= f^{367} = - \dfrac{1}{2} \,, \\
f^{458} &= f^{678} = \dfrac{\sqrt{3}}{2} \,,
\label{eq:f-tensors}
\end{split}
\end{gather}
and their anti-symmetric counterparts. The non-vanishing symmetric constants are
\begin{gather}
\begin{split}
d^{118} &= d^{228} = d^{338} =  \dfrac{1}{\sqrt{3}} \,, \\
d^{888} &= -\dfrac{1}{\sqrt{3}} \,, \\
d^{448} &= d^{558} = d^{668} = d^{778} = -\dfrac{1}{2\sqrt{3}} \,, \\
d^{146} &= d^{157} = d^{256} = d^{344} = d^{355} = \dfrac{1}{2} \,, \\
d^{247} &= d^{366} = d^{377} = - \dfrac{1}{2} \,,
\label{eq:d-tensors}
\end{split}
\end{gather}
and their symmetric counterparts.

Two useful properties are straightforward consequences of Eqs.~(\ref{eq:f-tensors}) and (\ref{eq:d-tensors}):
\begin{itemize}
\item Property A: For any pair of indices $a,b$, there is at most one value of a third index $c$ for which $d^{abc} \neq 0$.
\item Property B: There is no combination of flavour indices $a,b,c$ for which $f^{abc} \neq 0$ and $d^{abc} \neq 0$. In other words, when $f^{abc} \neq 0$, then $d^{abc} = 0$, and when $d^{abc} \neq 0$, then $f^{abc} = 0$.
\end{itemize}

	\section{Renormalisation and improvement}
\label{app:renimp}

All operators of interest are flavour non-singlets and, unless otherwise stated, quark masses are degenerate. For Wilson fermions, with $\rmO(a)$ Symanzik improvement, we know that the improved current
\begin{eqnarray}
\label{eq:Aimpr}
(A_{\rm I})^a_\mu  = A^a_\mu \, + \, a c_{\rm A} \partial_\mu P^a \,,
\end{eqnarray}
is correctly normalised  a follows:
\begin{eqnarray}
\label{eq:A-imp}
(A_{\rm R})^a_\mu  = Z_\mathrm{A} \, [1 + b_\mathrm{A} a m_\mathrm{q} + \bar b_\mathrm{A} a \Tr M_\mathrm{q}]  (A_{\rm I})^a_\mu \,.
\end{eqnarray}
The renormalised and Symanzik-improved scalar and pseudoscalar densities are given by
\begin{align}
S_{\rm R}^a  &= Z_\mathrm{S}  [1 + b_\mathrm{S} a m_\mathrm{q} + \bar b_\mathrm{S} a \Tr M_\mathrm{q} ] S^a \,,
\label{eq:S-imp}
\\
P_{\rm R}^a &= Z_\mathrm{P}  [1 + b_\mathrm{P} a m_\mathrm{q} + \bar b_\mathrm{P} a \Tr M_\mathrm{q} ] P^a \,,
\label{eq:P-imp}
\end{align}
with $a m_\mathrm{q} = 1/(2\kappa) - 1/(2 \kappa_{\mathrm{crit}})$ the subtracted bare mass; here
$\kappa$ is the Wilson hopping parameter and $\kappa_{\mathrm{crit}}$ its critical value (chiral limit).
The mass matrix of subtracted quark masses is denoted by $M_\mathrm{q}$.
The current (bare) quark mass, which appears in the chiral Ward identities of the present paper, is defined by the PCAC relation
\begin{align}
\label{eq:PCACmass}
m  =  \dfrac{\partial_0 \langle (A_{\rm I})^a_0(x)  \,\, \cO^a \rangle}{2 \, \langle P^a(x) \, \cO^a \rangle} \,.
\end{align}
The renormalised quark mass $m_{\mathrm R}$ is given in terms of the current mass $m$ by
\begin{align}
\label{eq:renmass-PCAC}
m_{\mathrm R}  =  \dfrac{Z_\mathrm{A}}{Z_\mathrm{P}} \,\, \dfrac{[1 + b_\mathrm{A} a m_\mathrm{q} + \bar b_\mathrm{A} a \Tr M_\mathrm{q}]}{[1 + b_\mathrm{P} a m_\mathrm{q} + \bar b_\mathrm{P} a \Tr M_\mathrm{q}] } \,\, m \,.
\end{align}

For two distinct flavours $i,j$, the subtracted quark masses are $a m_{{\mathrm q},i} = 1/(2\kappa_i) - 1/(2 \kappa_{\mathrm{crit}})$ and similarly for $a m_{{\mathrm q},j}$. The PCAC mass is defined as
\begin{align}
\label{eq:PCACmassij}
m_{ij}  =  \dfrac{\partial_0 \langle (A_{\rm I})^{ij}_0(x)  \,\, \cO^{ji}\rangle}{2 \, \langle P^{ij}(x) \,  \cO^{ji} \rangle} \,,
\end{align}
and the renormalised quark mass average is expressed in terms of $m_{ij}$ as follows:
\begin{align}
\begin{split}
\dfrac{m_{i,\rm R} + m_{j,\rm R}}{2}  & =  \dfrac{Z_\mathrm{A}}{Z_\mathrm{P}} \,\, m_{ij} \,\,
\Big [ 1 \, + \, (b_\mathrm{A}-b_\mathrm{P}) a m_{\mathrm{q},ij} \\
& + \, (\bar b_\mathrm{A}- \bar b_\mathrm{P})a \,\Tr M_\mathrm{q} \Big ]  + \rmO(a^2) \,,
\label{eq:renmassPCAC}
\end{split}
\end{align}
where $m_{\mathrm{q},ij} \equiv (m_{\mathrm{q},i} + m_{\mathrm{q},j})/2$. This reduces to Eq.~(\ref{eq:renmass-PCAC}) for two degenerate masses $m_{\mathrm{q},i} = m_{\mathrm{q},j}$.

In practice for the divergence of the improved axial current we use $\partial_\mu (A_{\rm I})^a_\mu \equiv
\tilde \partial_\mu A^a_\mu + a c_{\rm A} \partial_\mu^\ast \partial_\mu P^a $, where $\tilde \partial_\mu$ denotes the
average of the usual forward and backward derivatives defined as $a \partial_\mu f(x) \equiv f(x+a\hat\mu) - f(x)$ and
$a \partial_\mu^\ast f(x) \equiv f(x) - f(x-a\hat\mu)$.

	\section{Charge conjugation, \texorpdfstring{$\gamma_5$}{Gamma5}-Hermitici\-ty, and correlation functions}
\label{app:corr-funct-symm}
Wilson quark propagators in lattices with Schr\"odinger functional boundary conditions, on a fixed background gauge field, are
standard ones, denoted as $[\psi(y) \, \bar \psi(x)]_\mathrm{F}$, or
boundary-to-bulk ones like $[\zeta({\bf v}) \bar \psi(x)]_{\mathrm F}$\footnote{See Ref.~\cite{Luscher:1996vw} for their definitions.}. They all obey the $\gamma_5$-Hermiticity property; e.g.
\begin{align}
\begin{split}
[\psi(x) \, \bar \psi(y)]_\mathrm{F}^\dagger & =  \gamma_5 \,\, [\psi(y) \, \bar \psi(x)]_\mathrm{F} \,\, \gamma_5 \,, \\
[\zeta({\bf v}) \, \bar \psi(x)]_\mathrm{F}^\dagger  & = \gamma_5 \,\, [\psi(x) \bar \zeta({\bf v})]_\mathrm{F} \,\, \gamma_5 \,.
\end{split}
\label{eq:gamma5H}
\end{align}
Under charge conjugation\footnote{The Dirac matrix conventions used in the present work are those of Appendix~A of Ref.~\cite{Luscher:1996sc}. The charge conjugation conventions are those of Appendix~B of the same reference.}, the quark bilinear operators of interest transform as follows:
\begin{align}
\begin{split}
\bar \psi(x) T^a \gamma_5 \psi(y) & \rightarrow   \bar \psi(y) [T^a]^T \gamma_5 \psi(x) \,, \\
\bar \psi(x) T^a \gamma_0 \gamma_5 \psi(y) & \rightarrow  \bar \psi(y) [T^a]^T  \gamma_0 \gamma_5  \psi(x) \,,
\end{split}
\label{eq:charheconjg}
\end{align}
with $[T^a]^T$ the transpose of $[T^a]$. The time-boundary operators $\bar \zeta({\bf u}) \gamma_5 T^a \zeta({\bf v})$ and $\bar \zeta^\prime({\bf u}^\prime) \gamma_5 T^a \zeta^\prime({\bf v}^\prime)$ satisfy analogous properties. Note that in Eqs.~(\ref{eq:gamma5H}), Wick-contracted fermion fields are same-flavour functions, while in Eqs.~(\ref{eq:charheconjg}) they are vectors in flavour space.

We now concentrate on the r.h.s. of WI~(\ref{eq:SPImassintlatt}), and in particular on Eq.~(\ref{eq:OSO=FafaFbfb}) and the traces $F_{\mathrm{S;1}}$ and $F_{\mathrm{S;2}}$ of Table~\ref{tab:trBOOB}. Using the $\gamma_5$-Hermiticity properties of Eqs.~(\ref{eq:gamma5H}), it can be easily shown that $F_{\mathrm{S};2}(y_0) = F_{\mathrm{S};1}(y_0)^\dagger$. On the other hand, the traces of three flavour matrices $T^{dea}$ and $T^{aed}$ are given by Eq.~(\ref{eq:trTTT}). Putting everything together, the r.h.s. of the Ward identity~(\ref{eq:SPImassintlatt}) becomes
\begin{align}
\begin{split}
\label{eq:WIrhs-fSa}
\mathrm{WI~r.h.s.} = - \frac{a^{15}}{2} Z_{\mathrm S}  d^{bce} \Big [ & d^{ade} \Re \big \{F_{\mathrm{S};1}(y_0) \big \} \\
+ \mathrm{i} & f^{ade} \Im \big \{ F_{\mathrm{S};1}(y_0) \big \} \Big ] \,.
\end{split}
\end{align}
Next we apply charge conjugation to the correlation function $\langle \cO^{\prime a} \,\, S^e(y) \,\,\cO^d \rangle$. We see from Eq.~(\ref{eq:charheconjg}) that the transformation only affects the flavour matrices; instead of $\Tr ( T^a T^e T^d)$ we have $\Tr( T^{aT} T^{eT} T^{dT}) = \Tr( T^d T^e T^a)$ and instead of  $\Tr( T^d T^e T^a)$ we have $\Tr( T^{dT} T^{eT} T^{aT}) = \Tr( T^a T^e T^d)$. Thus, under a charge conjugation transformation,
 \begin{align}
 \begin{split}
\mathrm{WI~r.h.s.} \rightarrow - \frac{a^{15}}{2} Z_{\mathrm S}  d^{bce} \Big [ & d^{ade} \Re \big \{F_{\mathrm{S};1}(y_0) \big \} \\
- \mathrm{i} & f^{ade} \Im \big \{ F_{\mathrm{S};1}(y_0) \big \} \Big ] \,.
\end{split}
\end{align}
This should be equal to the original expression~(\ref{eq:WIrhs-fSa}), because charge conjugation leaves QCD correlation functions unaffected. Comparing the last two equations we see that this can only be true if $ \Im \big \{ F_{\mathrm{S (1)}}(y_0) \big \}$ vanishes. This proves Eq.~(\ref{eq:WIrhs}).

Having shown that the r.h.s. of WI~(\ref{eq:SPImassintlatt}) is real, the l.h.s. must also be real. As a crosscheck we show this explicitly. The l.h.s. correlation function is 
given by Eq.~(\ref{eq:TkFAPk}), with the traces of flavour matrices given by Eqs.~(\ref{eq:T12})--(\ref{eq:T9}) and the $9$ terms $F_{\mathrm{AP};k}$ listed in Table~\ref{tab:trBOOB}. Taking the Hermitean conjugate of these terms we find that the one-boundary ones are related pairwise by complex conjugation,
\begin{align}
\nonumber
T_2^{abcd}F_{\mathrm{AP};2}(x_0,y_0)&=[T_1^{abcd}F_{\mathrm{AP};1}(x_0,y_0)]^\ast \,,\\
\label{eq:F4T4}
T_4^{abcd}F_{\mathrm{AP};4}(x_0,y_0)&=[T_3^{abcd}F_{\mathrm{AP};3}(x_0,y_0)]^\ast \,,\\
\nonumber
T_6^{abcd}F_{\mathrm{AP};6}(x_0,y_0)&=[T_5^{abcd}F_{\mathrm{AP};5}(x_0,y_0)]^\ast \,.
\end{align}
Hermitean conjugation also implies that the quark-disconnected contributions are real:
\begin{align}
\nonumber
T_7^{abcd}F_{\mathrm{AP};7}(x_0,y_0)&=[T_7^{abcd}F_{\mathrm{AP};7}(x_0,y_0)]^\ast \,,\\
\label{eq:F8f8}
T_8^{abcd}F_{\mathrm{AP};8}(x_0,y_0)&=[T_8^{abcd}F_{\mathrm{AP};8}(x_0,y_0)]^\ast \,,\\
\nonumber
T_9^{abcd}F_{\mathrm{AP};9}(x_0,y_0)&=[T_9^{abcd}F_{\mathrm{AP};9}(x_0,y_0)]^\ast \,.
\end{align}
From these properties it immediately follows that the l.h.s. of the WI is real.

However we want to go a step further and show the reality of the traces $F_{\mathrm{AP};1}, \ldots, F_{\mathrm{AP};9}$. For the one-boundary contributions, Eqs.~(\ref{eq:F4T4}) imply that
\begin{align}
&T_1^{abcd} F_{\mathrm{AP};1} + T_2^{abcd} F_{\mathrm{AP};2} \label{eq:F1f1F2f2}\\
=\,&T_1^{abcd} F_{\mathrm{AP};1} + (T_1^{abcd} F_{\mathrm{AP};1})^\ast 
= 2 \Re[T_1^{abcd} F_{\mathrm{AP};1}] \nonumber\\
=\,&2 [ \Re(T_1^{abcd}) \Re(F_{\mathrm{AP};1}) - \Im(T_1^{abcd}) \Im(F_{\mathrm{AP};1}) ] \,, \nonumber
\end{align}
with (cf. Eq.~(\ref{eq:trTTTT})):
\begin{align}
\Re(T_1^{abcd}) & = \dfrac{1}{4 \nf} \delta^{ab} \delta^{cd} \, +  \dfrac{1}{8}  [d^{abe} d^{cde} - f^{abe} f^{cde} ] \,\,,
\nonumber \\ 
\Im(T_1^{abcd}) & = \dfrac{1}{8}  [d^{abe} f^{cde} + f^{abe} d^{cde} ] \,.
\label{eq:tr4T}
\end{align}
Applying charge conjugation to the 4-point correlation function $\langle \mathcal{O}^{\prime a} \,A_0^b(x) \, P^c(y) \,\mathcal{O}^d\rangle$, we find that $F_{\mathrm{AP};1} \rightarrow F_{\mathrm{AP};1}$, $F_{\mathrm{AP};2} \rightarrow F_{\mathrm{AP};2}$, and $T_1^{abcd} \leftrightarrow T_2^{abcd}$. Thus under charge conjugation Eq.~(\ref{eq:F1f1F2f2}) transforms as follows:
\begin{align}
&T_1^{abcd} F_{\mathrm{AP};1} + T_2^{abcd} F_{\mathrm{AP};2} \\
\rightarrow \,&2 [ \Re(T_2^{abcd}) \Re(F_{\mathrm{AP};1}) - \Im(T_2^{abcd}) \Im(F_{\mathrm{AP};1}) ] \,.\nonumber
\end{align}
But applying Eq.~(\ref{eq:trTTTT}) to $T_2^{abcd}$ (cf. also Eq.~(\ref{eq:tr4T})) we see that $\Re(T_2^{abcd}) = \Re(T_1^{abcd})$ and $\Im(T_2^{abcd}) = -\Im(T_1^{abcd})$. Thus, under charge conjugation
\begin{align}
&T_1^{abcd} F_{\mathrm{AP};1} + T_2 ^{abcd} F_{\mathrm{AP};2} \\
\rightarrow \,&2 [ \Re(T_1^{abcd}) \Re(F_{\mathrm{AP};1}) + \Im(T_1^{abcd}) \Im(F_{\mathrm{AP};1}) ] \,. \nonumber
\end{align}
Comparing this result to Eq.~(\ref{eq:F1f1F2f2}) and recalling that QCD correlation functions remain invariant under charge conjugation, we deduce that $\Im(F_{\mathrm{AP};1})=0$. Analogously, $F_{\mathrm{AP};2}, \ldots , F_{\mathrm{AP};6}$ are also real.
Concerning one-boundary contributions, traces $T_7^{abcd}, T_8^{abcd} , T_9^{abcd}$ are easily seen to be real from Eq.~(\ref{eq:TTnorm}). The reality of $F_{\mathrm{AP};7}, F_{\mathrm{AP};8}, F_{\mathrm{AP};9}$ then follows immediately from Eqs. (\ref{eq:F8f8}). This completes our proof that also the l.h.s. of WI~(\ref{eq:SPImassintlatt}) is real.

	\section{Non-perturbative checks}
\label{app:np_identities}
\begin{figure}[h]
	\centering
	\includegraphics[width=\linewidth]{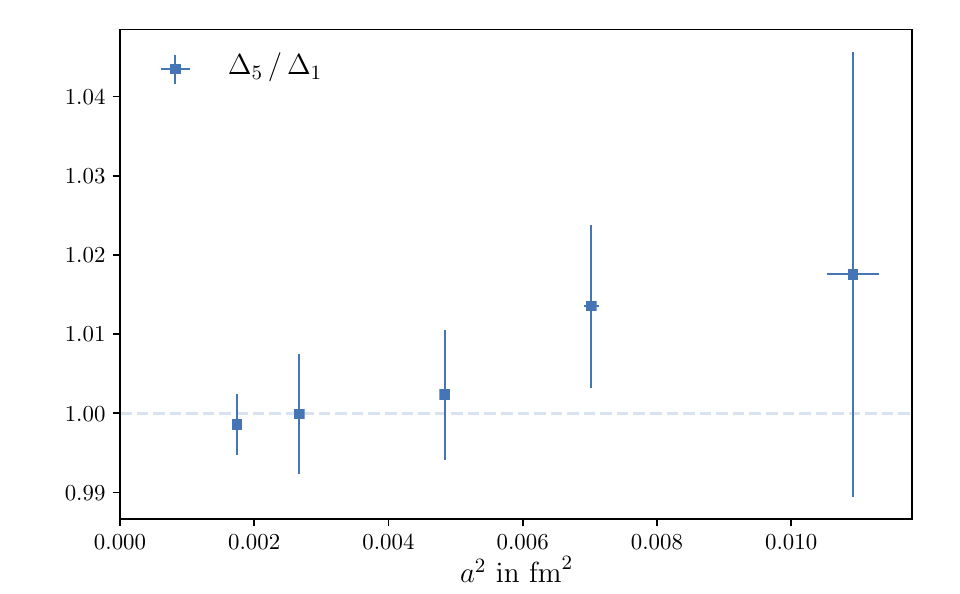}
	\includegraphics[width=\linewidth]{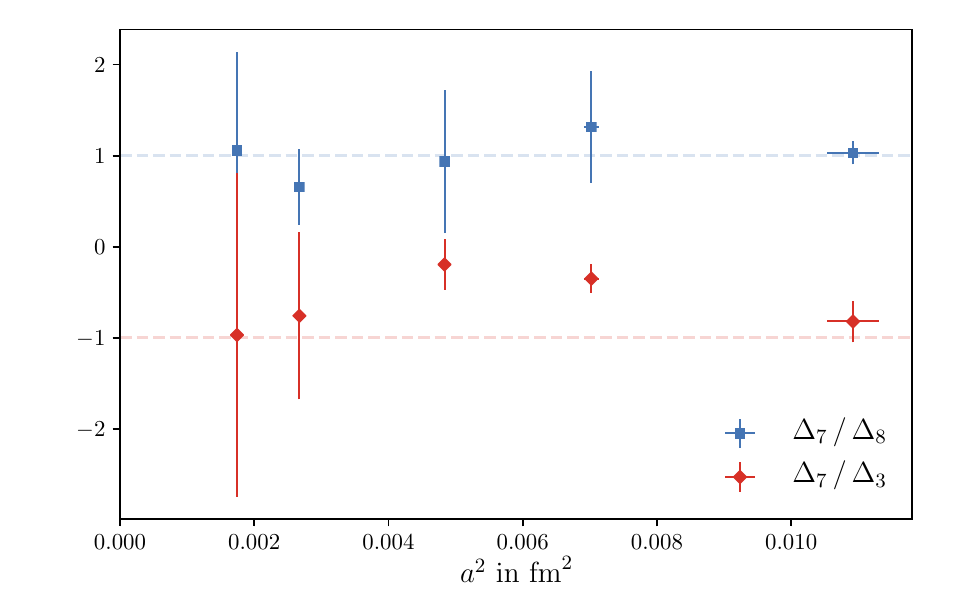}
	\caption{Non-perturbative confirmation of the identities (\ref{eq:D1=D5}), (\ref{eq:D7=D8}) and (\ref{eq:D3=-D7}). In the top panel only quark-connected diagrams contribute to the results, while in the bottom panel quark-disconnected diagrams also give contributions.}
	\label{fig:ratio_d5_d1}
\end{figure}

As additional validation of our method we want to make sure that the relations (\ref{eq:D1=D5}), (\ref{eq:D7=D8}) and (\ref{eq:D3=-D7}) which relate different diagrams to one another are fulfilled up to ambiguities of $\mathrm{O}(a^2)$.
After making sure that the identities are valid at tree-level of perturbation theory we evaluate them non-perturbatively on our ensembles. The analysis is analogous to the one for the ratio $Z_\mathrm{S}/Z_\mathrm{P}$. After evaluating the identities on each lattice for a given value of $\beta$, we perform an extra- or interpolation to the chiral point linear in the current quark mass. The values presented here are the results at the chiral point obtained from this procedure.
The clearest evidence comes from identity (\ref{eq:D1=D5}) which we can rewrite as
\begin{align}
\Delta_5\, / \, \Delta_1 &= \phantom{-} 1 + \mathrm{O}(a^2)\,.
\end{align}
In the top part of Fig. \ref{fig:ratio_d5_d1} we present the results which show the expected scaling towards the continuum. 
The identities (\ref{eq:D7=D8}) and (\ref{eq:D3=-D7}) are more complicated to verify as they involve quark disconnected contributions. We can rewrite the identities as follows
\begin{align}
\Delta_7\, / \, \Delta_8 &= \phantom{-} 1 + \mathrm{O}(a^2)\,,\\
\Delta_7\, / \, \Delta_3 &= -1 + \mathrm{O}(a^2)\,.
\end{align}
The numerical results are presented in the bottom part of Fig. \ref{fig:ratio_d5_d1}. In this case the statistical uncertainties are or\-ders of magnitudes larger and grow towards the con\-tinuum limit. A possible explanation of this is that the $\Delta_i$ involved here are vanishing at tree-level in per\-tur\-bation theory. Despite the large uncertainties our data still suggest that the identities are fulfilled up to the expected ambiguities in the lattice spacing.

\end{appendix}
\addcontentsline{toc}{section}{References}
\bibliographystyle{JHEP}
\bibliography{zszp_nf3}

\end{document}